\documentclass[conference, a4paper]{IEEEtran}
\IEEEoverridecommandlockouts

\ifCLASSINFOpdf
  \usepackage{graphicx}
  \DeclareGraphicsExtensions{.pdf,.jpeg,.png}
\else
\fi

\usepackage{cite}
\usepackage{amsmath,amssymb,amsfonts}
\usepackage{algorithm,algorithmic}
\usepackage{graphicx}
\usepackage{epstopdf}
\usepackage{textcomp}
\usepackage{xcolor}
\usepackage[left=1.4cm, right=1.4cm, top=1.7cm]{geometry}
\usepackage[caption=false]{subfig}
\usepackage{anyfontsize}
\usepackage{csquotes}
\usepackage{amsthm}
\usepackage{mathtools}
\usepackage{bm}
\usepackage{enumitem}
\usepackage{flushend}
\usepackage{bbm}
\usepackage{hyperref}
\usepackage{multirow}
\usepackage{cancel}
\usepackage{array}

\makeatletter
\let\oldlt\longtable
\let\endoldlt\endlongtable
\def\longtable{\@ifnextchar[\longtable@i \longtable@ii}
\def\longtable@i[#1]{\begin{figure}[t]
\onecolumn
\begin{minipage}{0.5\textwidth}
\oldlt[#1]
}
\def\longtable@ii{\begin{figure}[t]
\onecolumn
\begin{minipage}{0.5\textwidth}
\oldlt
}
\def\endlongtable{\endoldlt
\end{minipage}
\twocolumn
\end{figure}}
\makeatother

\DeclareSymbolFont{ugmL}{OMX}{mdugm}{m}{n}
\SetSymbolFont{ugmL}{bold}{OMX}{mdugm}{b}{n}
\DeclareMathAccent{\wideparen}{\mathord}{ugmL}{"F3}

\newtheorem{prop}{Proposition}
\newtheorem{asu}{Assumption}

\makeatletter
\newcommand*\titleheader[1]{\gdef\@titleheader{#1}}
\AtBeginDocument{%
\let\st@red@title\@title
\def\@title{%
\bgroup\normalfont\normalsize\centering\@titleheader\par\egroup
\vskip0.2em\st@red@title}
}
\makeatother

\makeatletter
\renewcommand{\fnum@figure}{Figure \thefigure}
\makeatother

\title{{Remote Identification Trajectory Coverage\\ in Urban Air Mobility Applications}\\
\thanks{\noindent\textsuperscript{$\star$}Authors contributed equally; \textsuperscript{$\dagger$}corresponding author. This research project was supported in part by NSF IUCRC Phase I: Center for Autonomous Air Mobility and Sensing (CAAMS) Award No. 2137195.}
}

\titleheader{}

\author{\IEEEauthorblockN{Hejun Huang\textsuperscript{1,$\star$}, Billy Mazotti\textsuperscript{1,$\star$}, Joseph Kim\textsuperscript{2}, Max Z. Li\textsuperscript{1,3,$\dagger$}}
\IEEEauthorblockA{\textsuperscript{1}Department of Aerospace Engineering\\
\textsuperscript{2}Department of Robotics\\
\textsuperscript{3}Department of Industrial and Operations Engineering\\
University of Michigan\\
Ann Arbor, MI, USA\\
\{\href{mailto:hejun@umich.edu}{hejun}, \href{mailto:bmazotti@umich.edu}{bmazotti}, \href{mailto:jthkim@umich.edu}{jthkim}, \href{mailto:maxzli@umich.edu}{maxzli}\}@umich.edu}
}

\renewcommand\IEEEkeywordsname{Keywords}
\IEEEaftertitletext{\vspace{-1\baselineskip}}

\begin{document}

\maketitle

% REMOVE PAGE NUMBERS BEFORE SUBMITTING -- ML JAN 2023 %%%%%%%%%%%%%%%
% \thispagestyle{plain}
% \pagestyle{plain}

\noindent \begin{abstract}
As Urban Air Mobility (UAM) and Advanced Air Mobility (AAM) continue to mature, a safety-critical system that will need to be implemented in tandem is Remote Identification (Remote ID) for uncrewed aircraft systems (UAS). To ensure successful and efficient deployment (e.g., maximal surveillance of UAS trajectories), as well as to better understand secondary impacts (e.g., consumer privacy risks in collecting real-time UAS trajectory information), the \emph{coverage} of broadcast-receive Remote ID architectures needs to be characterized. Motivated by this need, we examine theoretical and empirical trajectory coverage of several common Remote ID technologies (e.g., Bluetooth, Wi-Fi) deployed for urban package delivery missions, a commonly-cited use case for UAM and AAM. We derive methods to explicitly compute expected coverage proportions under idealized geometries, as well as conduct case studies with realistic city geographies and UAS path planning algorithms. An example of results include approximate magnitudes of Remote ID receivers needed (approximately 500-5000 receivers needed to achieve 50-95\% coverage for Bluetooth Legacy, and approximately 10-40 receivers needed for the same coverage range for Wi-Fi NAN/Beacon, assuming a cruise altitude of 200 feet) to achieve specific trajectory coverage proportions for San Francisco, California. Our analyses, combined with complementary works related to Remote ID bandwidth and deployment topologies, can help guide municipal authorities and AAM stakeholders in future Remote ID system deployments and upkeep.
\end{abstract}

%\vspace{0.3cm}


\section{Introduction}		\label{sec:intro}

Uncrewed aircraft systems (UAS), also referred to colloquially as drones, are anticipated to play a large role in the future of aerial mobility. There is a diverse range of UAS vehicle classes, corresponding to a wide array of use cases and mission applications that previously were infeasible (i.e., unserved) or underserved \cite{goyal2021_AAM}. This expansion of the air transportation system into new modalities and services is encompassed within Advanced Air Mobility (AAM) \cite{NASAAAM_2022} -- of particular interest is the subset of AAM focused on urban applications: Urban Air Mobility (UAM) is generally differentiated from other contexts such as Rural Air Mobility (RAM) in terms of use cases and constraints. UAM use cases include last mile-type, potentially even vendor-to-door package and food delivery services, passenger air taxi and shuttle operations \cite{cohen2021_AAMUseCases}, as well as rapid transport of time-sensitive goods such as donated organs and emergency medical equipment \cite{sigari2021_AAMM}. Constraints unique to UAM include collision risks with dense infrastructures, difficulties in forecasting urban weather and environmental conditions, and acceptance by large communities with different attitudes regarding and perceptions of UAM \cite{cohen2021_UC}.

Given the numerous safety and security concerns revolving around UAS and AAM applications, several safety-critical adjacent systems have been proposed \cite{ellis2021_IASMS}. One of these systems -- \emph{Remote Identification} (Remote ID) in the US \cite{FAA_RemoteID_2021,RID_FR} and analogous counterparts internationally (e.g., \cite{EASA_FR}) -- has reached technical, legislative, and regulatory maturity. Remote ID requires the broadcast of identifying information (e.g., drone ID), real-time position and velocities, as well as control station location from drone takeoff to shutdown \cite{FAA_RemoteID_2021,RID_FR}. Through data blocks retrieved via Remote ID, the \emph{trajectory} of a drone can be tracked in real time, in addition to being subject to post hoc analysis. For brevity, we will refer to all remote identification-type systems as Remote ID.

Remote ID was published to the US Federal Register in 2021, and enforcement of mandatory compliance is expected prior to 2025 \cite{FAA_RemoteID_2021,RID_FR}. Unsuccessful lawsuits challenging the propriety of Remote ID \cite{RID_Court} further demonstrate commitments to and momentum for implementation: The deciding factor in favor of Remote ID rests on its safety-critical nature; this is best summarized through the following court opinion striking down a challenge to Remote ID:

\begin{quote}
Drones are coming. Lots of them. They are fun and useful. But their ability to pry, spy, crash, and drop things poses real risks. Free-for-all drone use threatens air traffic, people and things on the ground, and even national security. [The US] Congress recognizes as much. It passed a law in 2016 requiring the Federal Aviation Administration (FAA) to \enquote{develop[] ... consensus standards for remotely identifying operators and owners of unmanned aircraft systems} and to \enquote{issue regulations or guidance, as appropriate, based on any standards developed.} \cite{RID_Court}
\end{quote}

The successful initiation and maturation of UAM (and AAM more broadly) will require safety-critical systems such as Remote ID to be implemented effectively (e.g., reliability in terms of drone tracking and reporting) and efficiently (e.g., without overbuilding required physical infrastructure such as Remote ID receivers). Furthermore, unintended negative externalities from such systems should be characterized in the context of real-world operations, with resultant mitigation strategies.

% - end

\subsection{Technical gap and research problem}  \label{ssec:problem_statement}

Even though Remote ID broadcast standards have been published (e.g., Bluetooth 4, Bluetooth 5, and Wi-Fi-based \cite{astm2022_remoteidstandard}), to our knowledge there has not been a rigorous examination of how \emph{comprehensive} Remote ID-type systems are for monitoring drone operations, particularly in UAM use cases. Motivated by the setting of UAM operations and focusing on Remote ID systems, in this work we conduct an analysis of Remote ID coverage in idealized settings and real-world case studies. Specifically, we are interested in determining what proportion of a given drone trajectory might be surveilled by a Remote ID receiver: Given a (fixed) Remote ID receiver with its coverage area, the trajectory of a drone flown entirely within its coverage area is considered 100\% covered, i.e., a coverage proportion of $1$.

In the idealized setting, we assume specific geometries for the Remote ID coverage area, in addition to how drone trajectories are generated: We derive analytical solutions to obtain the expected coverage proportion, and compare with Monte Carlo-based simulations to validate our solutions. We then conduct more realistic simulation-based case studies given geographic information of a major city, coverage properties (e.g., Bluetooth range), and different drone trajectory path planning techniques.

The results from our work can be used to inform how Remote ID infrastructure could be optimally deployed within an urban environment. It also provides a framework for analyzing how important factors such as the radius of the coverage area, Remote ID receiver distribution, and trajectory path planning method impact the effectiveness of Remote ID systems from the perspective of trajectory coverage. Finally, having better coverage -- as well as a better understanding of how the extent of the coverage changes with respect to aforementioned factors -- are critical inputs needed to mitigate negative externalities stemming from the Remote ID system itself (e.g., consumer privacy risks \cite{ding2022routing}) or from UAM operations (e.g., noise and visual nuisance \cite{woodcock2022_noise}, drone intrusion concerns \cite{wang2021_cuas}).

% risks from remote ID itself

% noise, etc

% why? use cases
% - mitigate privacy issues
% - efficiency + efficiacy

% A key safety-critical system 

\subsection{Related literature}  \label{ssec:previous_work}

% \cite{homola2017_utm,young2020_faatest}
Within the US, the final ruling on Remote Identification for uncrewed aircraft operating beyond visual line-of-sight (BVLOS) has been published to the US Federal Register \cite{FAA_RemoteID_2021,RID_FR}, with associated ASTM standards documented in \cite{astm2022_remoteidstandard}. Additionally, research development, including large-scale testing sites for the establishment of critical criteria such as Minimum Operational Performance Standards (MOPS) have been carried out \cite{young2020_faatest}.

% \cite{kato2019_400mhz,belwafi2022_ridts}
% \cite{wu2014_ltenetworks,zeng2018_cellular,tomaszewski2020_5g,abdalla2021_3gpp}
Two operating modalities for Remote ID are broadcast-based versus network-based. Broadcast-based Remote ID utilizes individual ground receiver modules which receive Remote ID broadcasts from active drones and UAS \cite{belwafi2022_ridts}. In addition to Bluetooth and Wi-Fi, long-range (LoRA) radio has also been explored for broadcast-based Remote ID \cite{mujumdar2021_lora}. Our work focuses on the broadcast-based modality, as network-based Remote ID operates in a fundamentally different manner \cite{wu2014_ltenetworks}. Our planar setup of ground station receivers and broadcast drone trajectories align with previous work in \cite{kuroda2020_uascoms} and \cite{mujumdar2021_lora}; for \cite{kuroda2020_uascoms}, an analysis of Bluetooth and Wi-Fi coverage in one hub-and-spoke network topology was performed using simulations. Our work complements and extends \cite{kuroda2020_uascoms} via an analytical derivation of expected coverage proportions, as well as realistic, random Remote ID receiver placements within a realistic UAM operating geography.

%\cite{sun2013_mobilesense,luo2019_tfe,tong2021_camera}
Finally, we note that the classical coverage problem, e.g., $k$-cover constructions wherein every point in an area is covered by at least $k$ sensors, has been well-studied \cite{megueridichian2001_coverage,huang2003_coverage}. However, these setups do not consider the coverage of a \emph{trajectory}, and mobile UAS have only been factored in as sensors themselves (e.g., \cite{paull2014_uavcoverage}), but not as the object of surveillance and coverage. More specifically, our setting can be considered as a trajectory-centered extension of the Boolean disc-type coverage set up; we refer interested readers to \cite{wang2011_coveragesurvey} for an extensive survey of Boolean disc-type coverage. Other related setups include the usage of fixed traffic sensors (e.g., stationary traffic cameras) to perform tasks such as congestion characterization, traffic flow estimation, and trajectory reconstruction \cite{luo2019_tfe,tong2021_camera}.

\section{Contributions of Work}		\label{sec:contributions}

The contributions of our work in this paper are as follows:

\begin{enumerate}
    \item We derive explicit formulas for expected coverage proportions under idealized environment and coverage area geometries. These expected coverage proportions depend on the method through which UAS trajectories are generated, as well as the Remote ID receiver technology under consideration. We validate these formulas through Monte Carlo simulations, and analyze differences in coverage proportions under different trajectory generation assumptions.
    \item We conduct a simulation-based case study centered on San Francisco, California, where we examine the number of Remote ID receivers required to attain specific trajectory coverage proportions (e.g., 50\% versus 95\% coverage) at different UAS cruising altitudes (200 feet versus 400 feet) and using different Remote ID receiver technologies (Bluetooth Legacy, Bluetooth Long Range, Wi-Fi NAN/Beacon).
    \item We provide an outline for a hybrid approach combining both the idealized analysis and realistic geographies. Additionally, we point out future directions for studying Remote ID deployment strategies in terms of broadcast-receiver architectures and coverage.
\end{enumerate}

% \mlcomment{TBC -- at very end}

% II. Contribution of Work % ML
%     - First analysis of Remote ID coverage for AAM applications
%     - Theory + Real world case studies
%     - Discuss (high-level) the results/insights
\section{Methods and Data}  \label{sec:methods_data}

We present an idealized analysis of the Remote ID coverage problem in Section \ref{ssec:ideal_anl}, and derive expressions for expected coverage proportions given assumptions on the environment geometry and origin-destination (OD) pair generation. We then describe the setup and data used in our urban area simulation-based case studies in Section \ref{ssec:urban_area_method_data}.

%To illustrate our work about the coverage proportion from formulation to reality, we propose an idealized mathematical statement in Section \ref{sec:ideal_anl}, a following real-world data configuration in Section \ref{sec:ideal_data}, which builds the core part of the discussions in \ref{sec:urban_cases}.

\subsection{Idealized analysis}\label{ssec:ideal_anl}

The setup for our idealized analysis of the Remote ID coverage problem is as follows: We define the \emph{coverage area} of the Remote ID receiver to be a disk $\mathcal{D}_c$ with radius $r_c > 0$ centered at $(x_0, y_0)$ in $\mathbb{R}^2$. We define the \emph{environment} containing possible OD pairs to be a circle $\mathcal{B}_e$ with radius $r_e \geq r_c > 0$, and we assume that the environment is centered at the same coordinate $(x_0, y_0)$ as $\mathcal{D}_C$. Future extensions of this analysis could generalize the centers of $\mathcal{D}_c$ and $\mathcal{B}_e$ such that they do not coincide.

\begin{figure}[!htbp] 
	\centering
	\includegraphics[scale=0.2]{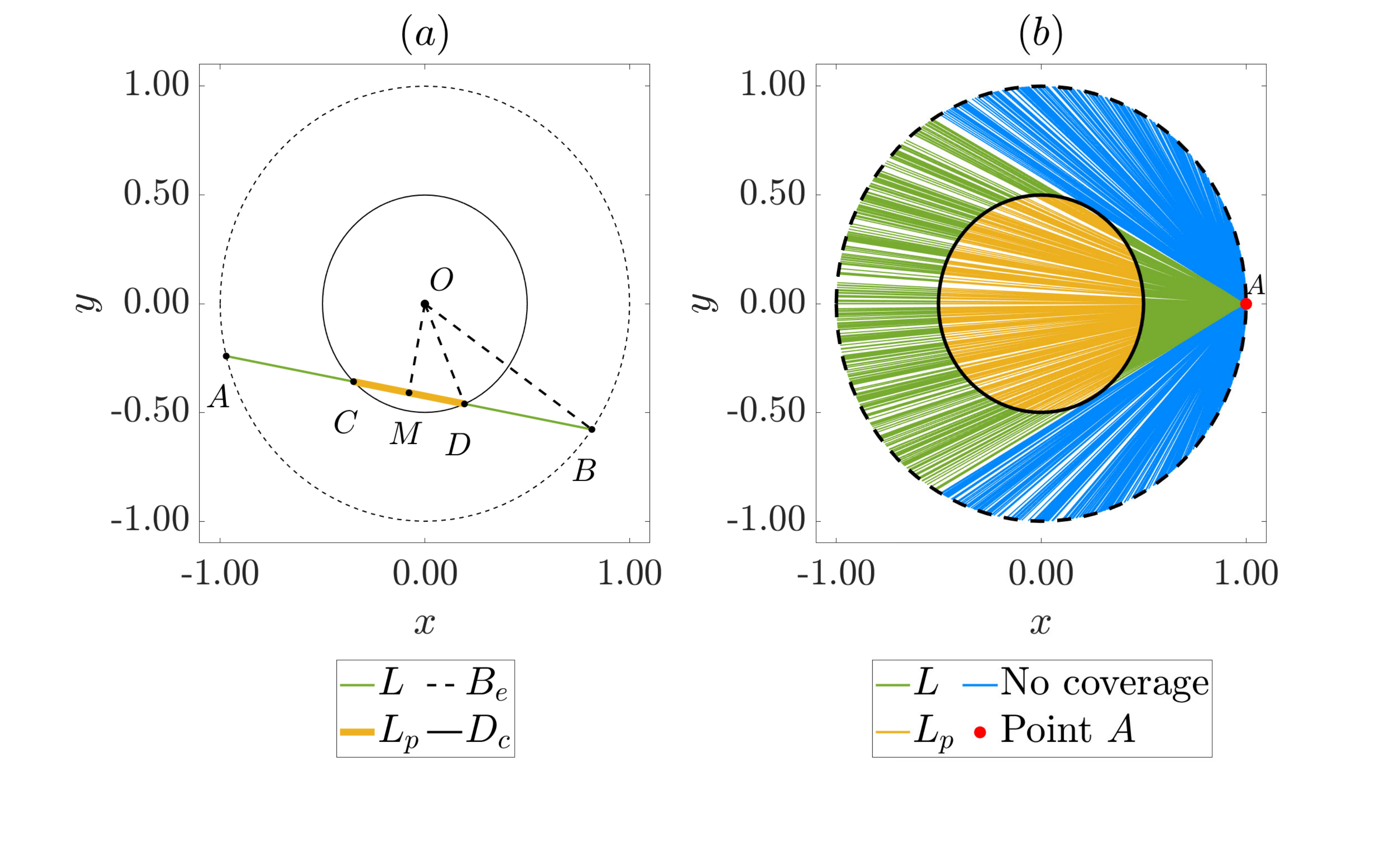}
	\caption{\emph{(a)} Remote ID receiver coverage area $\mathcal{D}_c$ within environment $\mathcal{B}_e$, with points of interest annotated; \emph{(b)} Fixing point $A$ with $\alpha = 0$.}
	\label{fig:prob_circ}
\end{figure}

To generate an OD pair, we select two points $A, B \in \mathcal{B}_e$ in the environment at random (to be formalized later in this section). Note that $A$ and $B$ lie on the circle $\mathcal{B}_e$, and can be described by the coordinates $\left(r_e\cos(\alpha), r_e\sin(\alpha) \right)$ and $\left(r_e\cos(\beta), r_e\sin(\beta) \right)$, respectively, for some angles $\alpha, \beta \in [0, 2\pi]$. We can describe the midpoint $M$ of the straight-line drone trajectory between OD pairs $A$ and $B$, which \emph{may or may not} lie in $\mathcal{D}_c$, by the coordinates

\begin{equation}
    \left( \frac{r_e(\cos(\alpha)+\cos(\beta))}{2},\frac{r_e(\sin(\alpha)+\sin(\beta))}{2} \right).
    \label{eq:mid}
\end{equation}

\noindent
Now, if $M$ lies in the coverage area $\mathcal{D}_c \setminus \partial\mathcal{D}_c$, we note that the straight-line trajectory intersects the boundary $\partial\mathcal{D}_c$ of the coverage area at two locations $D$ and $C$. In this case, we denote by $L$ and $L_p$ the portions of the straight-line trajectory outside and inside $\mathcal{D}_c$, respectively, and $\ell$ as the distance between the midpoint and $(x_0, y_0)$ (i.e., the length of the line segment $\overline{OM}$ in panel (\emph{a}) of Figure \ref{fig:prob_circ}). We have the following proposition:

\begin{prop}[Coverage proportion] Let $P \in (0, 1]$ be the \emph{coverage proportion}. If the straight-line trajectory intersects the boundary $\partial\mathcal{D}_c$ of the coverage area in two locations, we have that

\begin{equation}
    P = \frac{L_p}{L} = \sqrt{\frac{r_c^2 - \ell^2}{r_e^2 - \ell^2}}.
    \label{eq:exp_cov}
\end{equation}
\label{prop:cov_proportion}
\end{prop}

\noindent\emph{Proof.} Following notation in Figure \ref{fig:prob_circ}(\emph{a}), the ratio $L_p/L$ is the same as the ratio of the lengths of line segments $\overline{MD}$ and $\overline{MB}$ since $M$ is the midpoint. This latter ratio is the same as the ratio of the areas of triangles $\Delta_{OMD}$ and $\Delta_{OMB}$ as the two triangles share the same height $\ell$. The expression in \eqref{eq:exp_cov} follows after noting that the areas of $\Delta_{OMD}$ and $\Delta_{OMB}$ are $\ell\sqrt{r^2_c-\ell^2}/2$ and $\ell\sqrt{r^2_e-\ell^2}/2$, respectively.\hfill$\square$

We now return to the key word of \emph{random} when selecting OD pairs to generate the straight-line trajectory for which we are interested in the coverage proportion $P$. Given the reliance of Proposition \ref{prop:cov_proportion} on the midpoint $M$, we expect that, if we were to sample random straight-line trajectories, the distribution of $M$ in the coverage area will be important. However, generating this trajectory -- which is precisely a chord of $\mathcal{B}_e$ -- is ambiguous: This is known as the Bertrand paradox \cite{marinoff1994_bp}. For the purposes of our Remote ID coverage analysis, we are interested in two cases: (i) Uniformly distributed endpoints, and (ii) uniformly distributed midpoints.

% \begin{remark}\label{rem:1}
% Notice that each chord's midpoint to the center resulting in a fixed coverage proportion $P_{\cap}$. But the way to generate a chord on the circle is ambiguous, which is known as the \href{https://en.wikipedia.org/wiki/Bertrand_paradox_(probability)}{\textit{Bertrand Paradox}}. Therefore, a chord's selection strategy is different from \textbf{uniform distributed endpoint} and \textbf{uniform distributed midpoint}. A clearer idealized analysis with different assumptions is given in the following section.
% \end{remark} 

\subsubsection{Case (i): Uniformly distributed endpoints}   \label{sssec:ude}

This case hinges on the following assumption:

\begin{asu}\label{asp:1}
When randomly selecting a straight-line trajectory (chord), we proceed by selecting endpoints that are uniformly distributed along $\mathcal{B}_e$.
%The endpoints of a chord in the circle $\bigcirc_E$ are uniformly distributed along the circumference.
\end{asu}

\noindent
Recall from above our endpoints $A, B$ with coordinates $\left(r_e\cos(\alpha), r_e\sin(\alpha) \right)$ and $\left(r_e\cos(\beta), r_e\sin(\beta) \right)$, respectively. Under Assumption \ref{asp:1}, we note that this is equivalently to selecting angles $\alpha, \beta$ uniformly, i.e., $\alpha$ and $\beta$ are drawn identically and independently from $\mathrm{Unif}[0, 2\pi]$. By symmetry, we note that we could set one of the angles to be fixed arbitrarily (see Figure \ref{fig:prob_circ}(\emph{b}) for intuition); without loss of generality, we set $\alpha = 0$. The coordinates for midpoint $M$ can now be rewritten as $\left(r_e(1+\cos(\beta))/2, r_e\sin(\beta)/2\right)$. 

% \begin{equation}
%     \left( \frac{r_e(1+\cos(\beta))}{2},\frac{r_e\sin(\beta)}{2} \right).
%     \label{eq:mid_UDE}
% \end{equation}

% A chord can be selected with given $2$ endpoints. This assumption depicts that each chord can be chosen in exactly one way, regardless of whether it is a diameter. Generally speaking, each pair of Remote ID were selected equally. We can reformulate (\ref{eq:circAB}) with uniformly distributed angles $\alpha,\beta\sim\mathcal{U}(0,2\pi)$. Because of the uniformity, the result generated from a fixed $\alpha$ is equal to the case of the random selected $\alpha$.

% \begin{figure}[t] 
% 	\centering
% 	\includegraphics[width=1\linewidth]{figures/Rid_case1_alpha0.eps}
% 	\caption{Fixed Starting Point $A$ with $\alpha=0$ for Coverage Ratio Exploration}
% 	\label{fig:demo_circ}
% \end{figure}

% Let $\alpha=0$  the midpoint can be obtained from (\ref{eq:mid}) as 
% \begin{equation}
%     \begin{aligned}
%     M(\frac{R_e(1+\cos(\beta))}{2},\frac{R_e\sin(\beta)}{2}).
%     \end{aligned}
% \end{equation}
% \noindent

In Figure \ref{fig:prob_circ}(\emph{b}) we visualize (via green-colored trajectories with yellow-colored portions covered by the Remote ID receiver) the intuition that the straight-line trajectory will only have a non-zero coverage proportion for certain ranges of $\beta$. However, since $\beta$ is a random variable, the squared distance $\ell^2$ from the midpoint $M$ to $(x_0, y_0)$ is a derived random variable, since

\begin{equation} \label{eqn:case1_x}
\begin{aligned}
    \ell^2 &= \left(\frac{r_e(1+\cos(\beta))}{2}\right)^2+\left(\frac{r_e\sin(\beta)}{2}\right)^2\\
    &=\frac{r_e^2(1+\cos(\beta))}{2}.
    \end{aligned}
\end{equation}

\noindent
Denote by $P_\mathrm{UDE}$ the coverage proportion under Assumption \ref{asp:1}. We observe that $P_\mathrm{UDE}$ is also a derived random variable, as it is a function of $\ell^2$ from Proposition \ref{prop:cov_proportion}. Let $\mathbb{E}\left[P_\mathrm{UDE}\right]$ be the \emph{expected} coverage proportion under Assumption \ref{asp:1}, we have that

\begin{prop}[Case (i) expected coverage proportion] Define the constant (deterministic) ratio $\rho = r_c/r_e$. We have that

\begin{equation}
    \mathbb{E}\left[P_\mathrm{UDE}\right] = \frac{1}{2\pi} \int_{\pi-2 \arcsin(\rho)}^{\pi+2\arcsin(\rho)} \sqrt{ \frac{ r_c^2 - r_e^2 \gamma(b) }{ r_e^2 - r_e^2 \gamma(b) } } \, db,
    \label{eq:sol_case1}
\end{equation}

\noindent
with $\gamma(b) = \frac{1+\cos(b)}{2}$.
    
\end{prop}

% \noindent\emph{Proof sketch.} \mlcomment{We omit full proof for brevity, check in with HH about $r_c$ limits ...}
\noindent\emph{Proof.} We omit the full proof for brevity, and give a sketch of the proof: As shown in Figure \ref{fig:prob_circ}(\emph{b}), the expectation of $P$ can be derived from the expectation of the angles governing the uniformly distributed endpoints. The upper and lower limits of (\ref{eq:sol_case1}) denote the interval of $\beta$ where coverage occurs. \hfill$\square$

% As shown in Figure \ref{fig:demo_circ}, the coverage will only happen within certain range of $\beta$. We can obtain the midpoint's distance square as
% \begin{equation}\label{eqn:case1_x}
%  x^2 = (\frac{R_e(1+\cos(\beta))}{2})^2+(\frac{R_e\sin(\beta)}{2})^2=\frac{R_e^2(1+\cos(\beta))}{2}.
% \end{equation}

% Based on (\ref{eqn:P_cap}) and (\ref{eqn:case1_x}), we can obtain the expectation of $P_{\cap}$
% \begin{equation}
%     \begin{aligned}\label{eq:sol_case1}
%         & \quad\; E[P^{(1)}_{\cap}] = \frac{4\arcsin(\frac{R_d}{R_e})}{2\pi} E[\sqrt{\frac{R_d^2-x^2}{R_e^2-x^2}}]+ \\
%         & \qquad\qquad\qquad\; \frac{2\pi -4\arcsin(\frac{R_d}{R_e})}{2\pi} 0\\
%         &= \frac{4\arcsin(\frac{R_d}{R_e})}{2\pi}  \int_{\pi-2\arcsin(\frac{R_d}{R_e})}^{\pi+2\arcsin(\frac{R_d}{R_e})} \frac{\sqrt{\frac{R_d^2-R_e^2(1+\cos(\beta))/2}{R_e^2-R_e^2(1+\cos(\beta))/2}}}{4\arcsin(\frac{R_d}{R_e})} d \beta,\\
%         &\quad\;\beta\in[\pi-2\arcsin(\frac{R_d}{R_e}),\pi+2\arcsin(\frac{R_d}{R_e})],\\
%         & \quad\; R_d \in (\frac{\sqrt{2}R_e\sqrt{1+\cos(\beta)}}{2},R_e)=(0,R_e).\\
%     \end{aligned}
% \end{equation}

% \noindent
% When $\alpha=0$, only $\beta\in[\pi-2\arcsin(\frac{R_d}{R_e}),\pi+2\arcsin(\frac{R_d}{R_e})]$ will occur coverage, otherwise $P_{\cap}= 0$. 
% % When $R_e = 1, R_c = 0.5$, $E[P^{(1)}_{\cap}] \approx 0.13397$.

%\subsubsection{Case 2: Uniform Distributed Midpoints}

\subsubsection{Case (ii): Uniformly distributed midpoints}   \label{sssec:udm}

Similar to case (i), we begin by stating the following assumption:

\begin{asu}\label{asp:2}
When randomly selecting a straight-line trajectory (chord), we proceed by selecting its midpoint such that it is uniformly distributed in $\mathcal{D}_c$.
\end{asu}

% \begin{asu}\label{asp:2}
% The midpoints of a chord in the circle $\bigcirc_E$ are uniformly distributed.
% \end{asu}

\noindent
We note that, excluding the case of a circle's diameter, choosing the midpoint fixes a unique chord, i.e., straight-line trajectory. We can exclude the case where the midpoint falls precisely at $(x_0, y_0) \in \mathcal{D}_c$ as this happens with zero probability (measure zero). Denote by $\ell$ the distance of the randomly chosen midpoint to $(x_0, y_0)$. We note that $\ell$ is a random variable taking values in $[0, r_e]$, and can write down its cumulative density function $F_\ell(\ell^\star)$ explicitly:

\begin{equation}
    F_\ell(\ell^\star) = \mathrm{Pr}\left(\ell \leq \ell^\star\right) = \begin{dcases}
\frac{\pi(\ell^\star)^2}{\pi r_e^2}, & \ell^\star \in [0, r_e], \\
0, & \text{otherwise}.
\end{dcases}
    \label{eqm:case2CDF}
\end{equation}

\noindent
Accordingly, the probability density function $f_\ell(\ell^\star)$ is

\begin{equation}
    f_\ell(\ell^\star) = \frac{d}{d\ell^\star} F_\ell(\ell^\star) = \begin{dcases}
\frac{2\ell^\star}{r_e^2}, & \ell^\star \in [0, r_e], \\
0, & \text{otherwise}.
\end{dcases}
    \label{eq:exp_case2}
\end{equation}

Denote by $P_\mathrm{UDM}$ the coverage proportion under Assumption \ref{asp:2}. We observe that $P_\mathrm{UDM}$ is a derived random variable as it is a function of $\ell$ from Proposition \ref{prop:cov_proportion}. Let $\mathbb{E}\left[P_\mathrm{UDM}\right]$ be the \emph{expected} coverage proportion under Assumption \ref{asp:2}, we have that

\begin{prop}[Case (ii) expected coverage proportion] We have that

\begin{equation}
    \mathbb{E}\left[P_\mathrm{UDM}\right] = \int_0^{r_c} \frac{2l}{r^2_e} \sqrt{ \frac{ r_c^2 - l^2 }{ r_e^2 - l^2 } } \, dl,
    \label{eq:sol_case2}
\end{equation}

\noindent
with $r_c \in \left(0, r_e\right)$.
    
\end{prop}

% \noindent\emph{Proof sketch.} \mlcomment{We omit full proof for brevity, ...}
\noindent\emph{Proof.} We omit the full proof for brevity, and give a sketch of the proof: Since $P_\mathrm{UDM}$ is a function of the random variable $l^\star$, its expectation can be obtained directly from the definition for the expectation of a continuous random variable. \hfill$\square$

\begin{figure}[t] 
	\centering
	\includegraphics[width=0.75\linewidth]{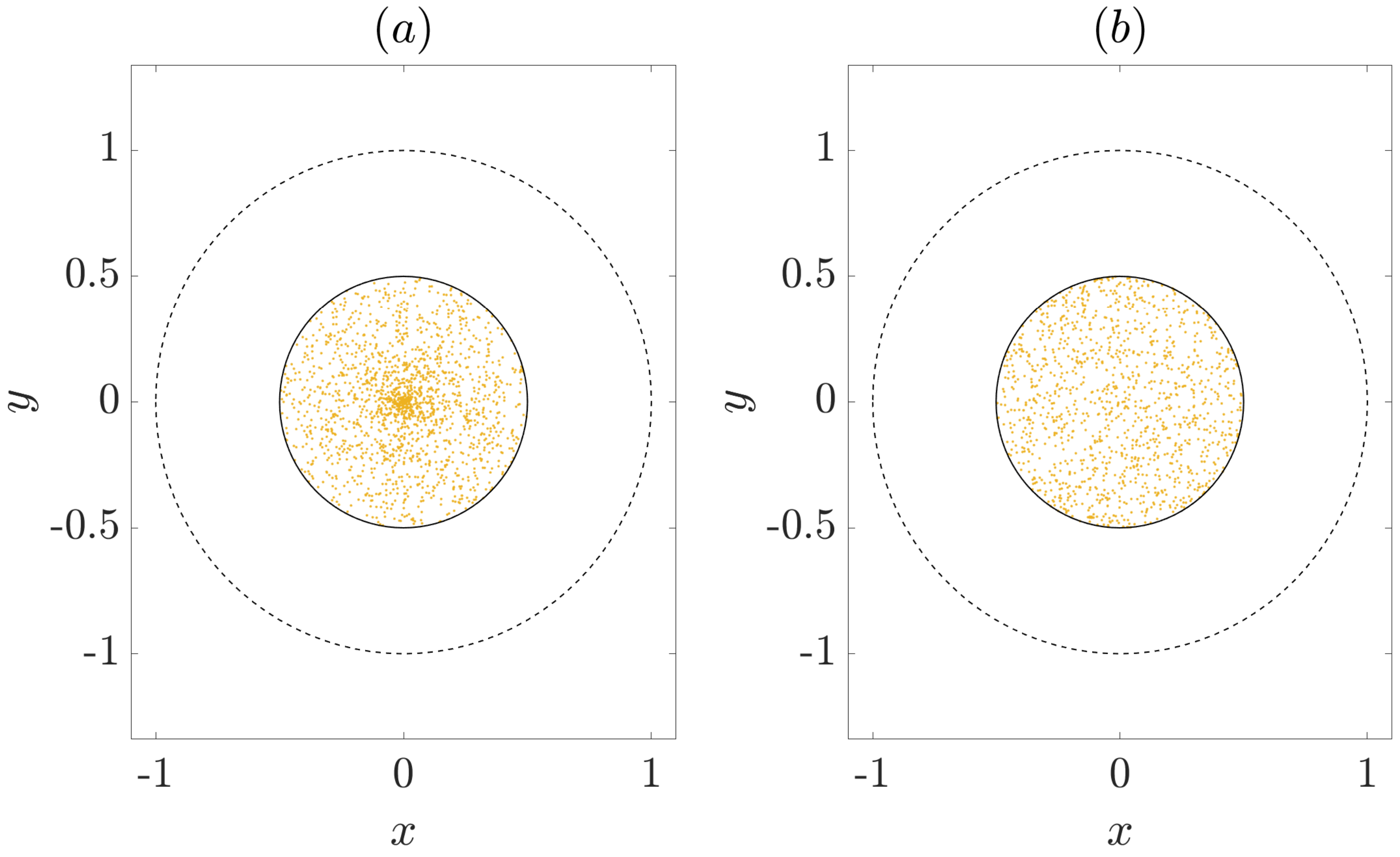}
	\caption{Midpoint distribution within $\mathcal{D}_c$ for (\emph{a}) Case (i) with uniformly distributed endpoints; (\emph{b}) Case (ii) with uniformly distributed midpoints.}
	\label{fig:midpoint_comp}
\end{figure}

The difference between the two expected coverage proportions in \eqref{eq:sol_case1} and \eqref{eq:sol_case2} is due to the difference in terms of midpoint distributions within the coverage area $\mathcal{D}_c$. Under the assumption of uniformly distributed endpoints (Assumption \ref{asp:1}), the resultant midpoint distribution is denser closer to $(x_0, y_0)$. By comparison, the midpoint distribution by definition under Assumption \ref{asp:2} is uniformly distributed in $\mathcal{D}_c$. We plot and confirm this graphically in Figure \ref{fig:midpoint_comp} with $R_e = 1$ and $R_c = 0.5$. Evaluating \eqref{eq:sol_case1} and \eqref{eq:sol_case2} numerically with $R_e = 1$ and $R_c = 0.5$ gives $\mathbb{E}\left[P_\mathrm{UDE}\right] \approx 0.134$ or $13.4$\% and $\mathbb{E}\left[P_\mathrm{UDM}\right] \approx 0.088$ or $8.8$\%. 

To summarize, we examined the Remote ID coverage problem under specific geometric assumptions of the coverage area and environment, given a Remote ID receiver with coverage radius $r_c$. We showed how differences in random trajectory generation can give rise to two different expected coverage proportions $\mathbb{E}\left[P_\mathrm{UDE}\right]$ and $\mathbb{E}\left[P_\mathrm{UDM}\right]$. Without using large-scale simulations, Remote ID location planning could be approximated as scaled-up versions of the idealized setup, and an estimation of expected coverage proportions made based off of, e.g., \eqref{eq:sol_case2}. In Section \ref{sec:ideal_coverage} we verify our expressions for $\mathbb{E}\left[P_\mathrm{UDE}\right]$ and $\mathbb{E}\left[P_\mathrm{UDM}\right]$ via Monte Carlo simulations, and also examine the difference $\Delta\mathbb{E}[P] = \mathbb{E}\left[P_\mathrm{UDE}\right] - \mathbb{E}\left[P_\mathrm{UDM}\right]$ in expected coverage proportions under Assumption \ref{asp:1} versus Assumption \ref{asp:2}.

% As we introduce in Remark \ref{rem:1}, the differences between (\ref{eq:sol_case1}) and (\ref{eq:sol_case2}) are caused by the midpoints' distribution. As shown in Figure \ref{fig:midpoint_comp} with $R_e=1,R_c=0.5$, we obtain $E[P^{(1)}_{\cap}] \approx 0.13397$ and $E[P^{(2)}_{\cap}] \approx 0.0884$, respectively. Based on Assumption \ref{asp:1} when coverage occurs, the midpoint's distribution in the left scheme is dense at the origin, while the midpoints in the right scheme are uniformly distributed. To sum up, Assumption \ref{asp:1} is more realistic and reasonable in Remote ID broadcast exploration.

% \subsubsection{Fixed OD Setup}\label{ssubsec:fixedOD}
% Compared with the idealized analysis, there are finite $n< \infty$ vertiports in the realistic scenario. Considered the construction cost, they can be regarded as invariant on the environment, such that there always exists or not exist combinations between any two vertiports. Without consider self-loop, these are $2(n-1),n\geq2,n\in\mathbb{Z}^+$ possible connections starting from one vertex and there are at most
% \begin{equation}\label{eqn:fullconnection}
% C =  2^{(n-1)} \cdot n,   
% \end{equation}
% \noindent
% non-repeating possible flight graphs in the environment. It's time-consuming, but we still can compute the coverage ratio between these finite vertiports. In Section \ref{sec:ideal_coverage} we will demonstrate the corresponding Monte-Carlo simulation of finite OD case and verify the idealized analysis (\ref{eq:sol_case1}) and (\ref{eq:sol_case2}) above. 

\subsection{Urban area simulations}\label{ssec:urban_area_method_data}

%\cite{leonard2021survey,kim2022airspace,ochoa2022urban,hohmann2022multi,malandrino2019planning,saxena2019optimal,wooten2019economic,jayaraman2022rooftop}
To simulate how choices in Remote ID receiver technologies, geographies, customer and vendor distributions, as well as path planning algorithms may affect overall trajectory coverage in real-world UAM settings, we use publicly available data sets for San Francisco, California, US. Additionally, for future studies, we prepared customer and vendor data sets for New York City and Los Angeles as well: We chose these cities based on their popularity in recent literature examining various UAM applications such as drone package delivery operations \cite{leonard2021survey,kim2022airspace,wooten2019economic,jayaraman2022rooftop}. In addition, these major population centers within the US have different densities of potential customers, vendors, and building structures (e.g., building heights), creating an ideal environment for our urban area simulations.

For the simulation environment, we set the origin and destination pair (OD pair) for one-way UAS flights to be located at the geolocations of real-world stores and residents. We assume that Remote ID receivers can be deployed on top of buildings, and that the centers of city building footprints serve as possible geolocations for these receivers. We retrieve the geolocation data sets used for this study from OpenStreetMap \cite{haklay2008openstreetmap}. Specific geolocation data are organized by MyGeoData \cite{mygeodata,orlik2014current} into \emph{themed} data sets. We chose to represent customer locations using MyGeoData's \enquote{Residential Land Use} theme; this theme provides general areas within a city that are predominantly occupied by single houses, grouped dwellings, apartments, flats, and units. We used MyGeoData's \enquote{Shopping Centers and Department Stores} themed data set, which provides building footprints and geolocation points of general stores, department stores, malls, supermarkets, and kiosks. Lastly, we chose MyGeoData's \enquote{Buildings} theme which provided the footprints and heights of individual and connected buildings. These building footprints and associated heights represent obstacles that must be avoided by the planned path of a drone. Additionally, these building footprints from the Building-theme data set serve as potential Remote ID receiver sites.%These building footprints and heights create the obsta, to create obstacle maps and receiver locations. %MyeGoData provides several additional themes covering  geolocations for real-world customers, vendors, and buildings, however due to limited resources we decided not to implement these data sets for this study.

\begin{table}[!htbp]
  \centering
  \caption{Summary of Remote ID receiver technologies}
    \begin{tabular}{ccc}
    \hline
    \multirow{1}{*}{Case Study Group} & \multicolumn{1}{c}{Radius}     &  \multicolumn{1}{c}{Technology} \\
    \hline
 % \multirow{1}{*}{\textbf{R250}} & 250 m &Bluetooth Legacy Advertising\\
 \multirow{1}{*}{\textbf{R250}} & 250 m &Bluetooth Legacy \cite{ASD-STAN_EuropeanRemoteID} \\
 \multirow{1}{*}{\textbf{R1000}} & 1 km &Bluetooth Long Range \cite{ASD-STAN_EuropeanRemoteID}\\
\multirow{1}{*}{\textbf{R2000}} & 2 km &Wi-Fi NAN, Wi-Fi Beacon \cite{ASD-STAN_EuropeanRemoteID}\\
\multirow{1}{*}{\textbf{Not Studied}} & Several km's & LoRa \cite{mujumdar2021_lora} \\
    \hline
     % \multicolumn{6}{c}{PE: Path Efficiency\;\; RT: Run Time}\\
     % \hline
    \end{tabular}%
  \label{tab:1_sheet1}%
\end{table}

For the urban case study experiments, we select between different coverage radii depending on the Remote ID receiver technology. This is equivalent to setting the $r_c$ parameter in the idealized analysis. Future work will involve the rigorous approximation of real-world geographies via the idealized analysis, e.g., exploring set partitioning of a city into simpler environments. In addition to varying the coverage radius of a given Remote ID receiver, we also vary the following for our experiments:

\begin{itemize}
    \item \emph{Path planning method.} We compare between two path planning approaches commonly used in UAS traffic management research. The first is simple straight-line path planning (\textsc{Slpp}) between the origin and the destination (used in, e.g., \cite{chin2021_utm}). We also explore rapidly exploring random trees (RRT\textsuperscript{*}) as a path planning algorithm \cite{karaman2011_pp}, which have been used in the context of UAM in, e.g., \cite{zammit2018_rrt}.
    \item \emph{Cruise altitude.} We explore two different cruising altitude for UAS, at 200 feet and at 400 feet.
    \item \emph{Coverage proportion.} This parameter is equivalent to $P, P_\mathrm{UDE},$ and $P_\mathrm{UDM}$ defined in the idealized analysis. For example, an average coverage proportion of 50\% indicates that, on average, 50\% of a given trajectory (with a fixed path planning algorithm) was covered by a Remote ID receiver.
\end{itemize}

\noindent
Additionally, we can explore different city geographies, given the appropriate base maps (e.g., customer and vendor distributions). We note that the altitudes we use are below current altitude maximums \cite{FAA2021_RSKY}; however, at further, lower altitudes, computation time becomes more significant, as more buildings and obstacles must be considered. %We can also explore additional desired coverage proportions, Remote ID radii, and geographies.

%  we can explore Remote ID coverage at different cruise altitudes, although we fix our experiments assuming a cruise altitude of 200 feet, which is below current altitude maximums

% We conduct different 

% MyGeoData's "Buildings" theme, providing the footprints and heights of individual and connected buildings, is used in creating obstacles for 

% Choices in coverage radii... What developed technologies have been seriously considered? (EuropeRemoteID \cite{XXX}, etc) -- Create Table (communication technology - range: Bluetooth Legacy Advertising - 250m, Bluetooth Long Range - 1km, Wi-Fi NAN - 2km, Wi-Fi Beacon - 2km, ); these will be the receiver ranges we use for the case studies. Other communication technologies include Name1,..., NameN with ranges R1,...,RN respectively.

% Choices in path planning algorithms... We further compare the effects of three different path planning approaches: origin-destination (OD), visibility graph (VG), and Rapidly-exploring Random Tree Star (RRT*). OD flight paths consist of a vertical takeoff and landing with shortest path cruise

% Choices in cities...

% Using OpenStreenMap data sets \cite{XXX} extracted by the MyGeoData geographical information system (GIS) converter provider \cite{XXX}, we obtain latitude and longitude coordinates for residential areas, shopping centers, and buildings.

% Recent path planning algorithms -- Three path planning approaches are in

% \mlcomment{TBC}

\section{Idealized Analysis Results}  \label{sec:ideal_coverage}

We first verify our expressions for the expected coverage proportions under Assumptions \ref{asp:1} and \ref{asp:2} using Monte Carlo simulations, where we randomly sample straight-line trajectories with OD pairs lying in the environment $\mathcal{B}_e$. We then examine the numerical differences between the expected coverage proportions under Assumption \ref{asp:1} versus \ref{asp:2}.

\subsection{Monte Carlo-based verification} \label{ssec:MC_verify}

\begin{figure}[!htbp]
  \centering
  \begin{tabular}{@{}c@{}}
    \includegraphics[width=1\linewidth]{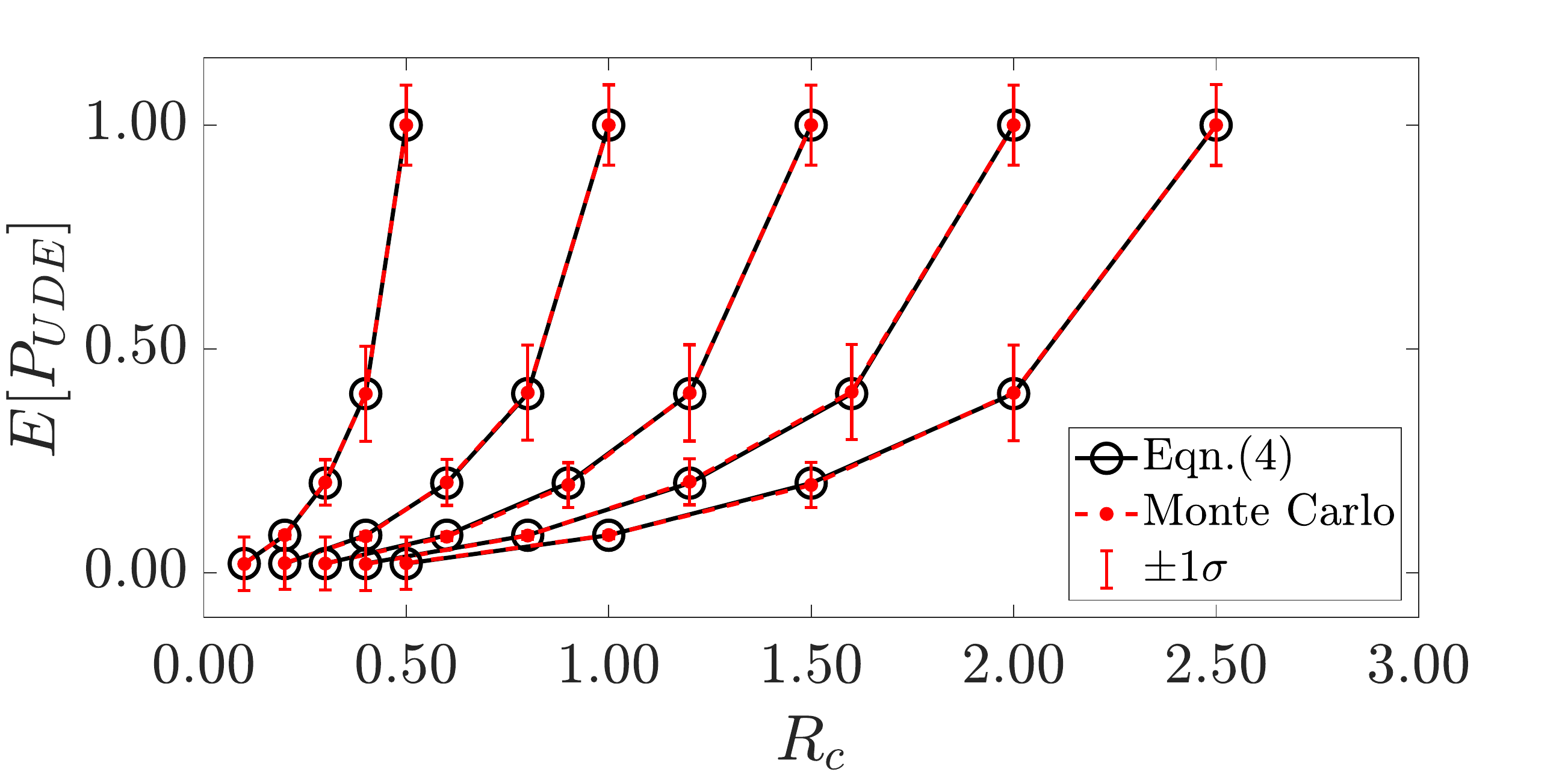} \\[\abovecaptionskip]
    \small \emph{(a)} Case (i)
  \end{tabular}

  \vspace{\floatsep}

  \begin{tabular}{@{}c@{}}
    \includegraphics[width=1\linewidth]{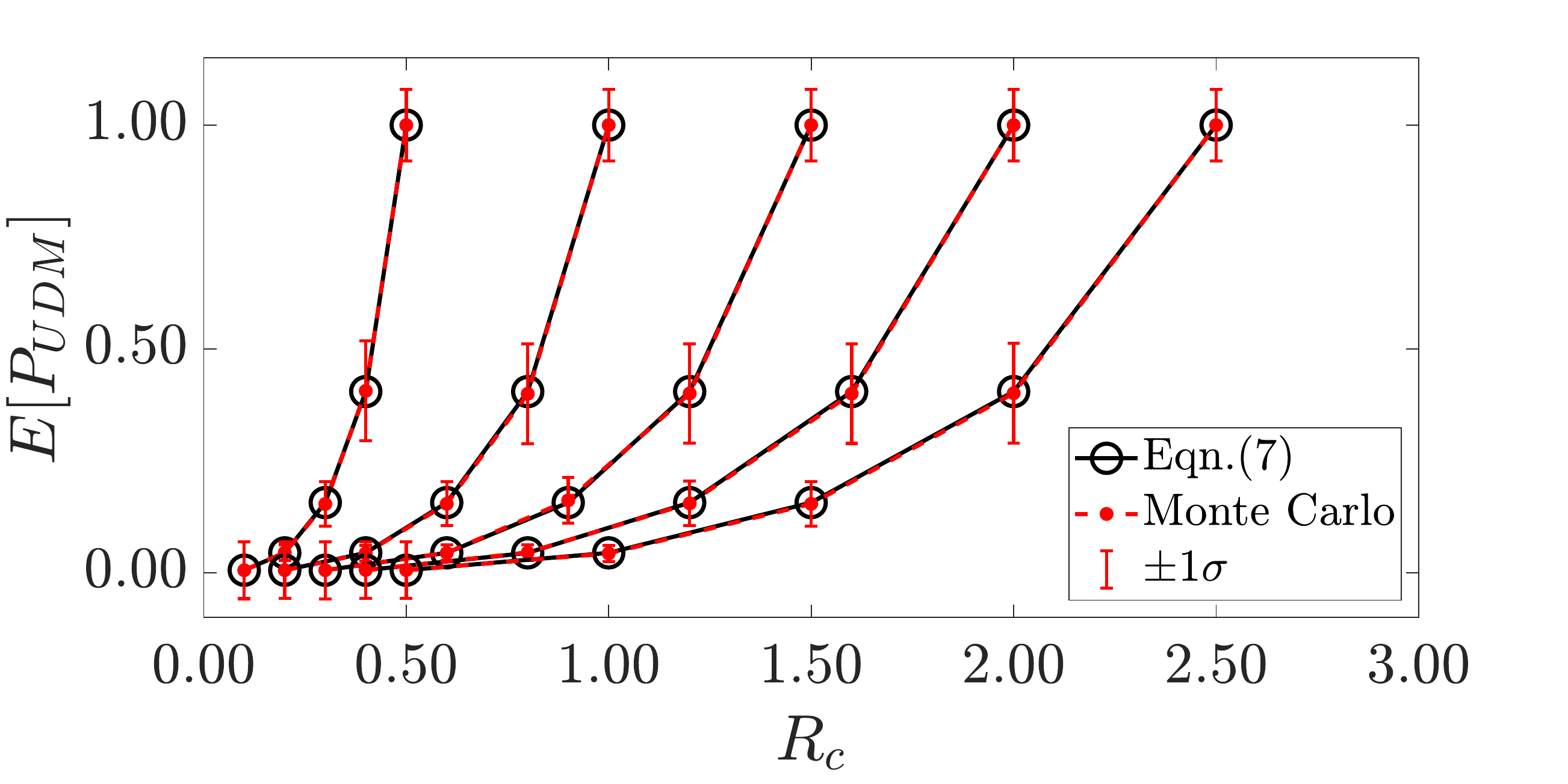} \\[\abovecaptionskip]
    \small \emph{(b)} Case (ii)
  \end{tabular}

    \caption{Comparisons between analytical expressions for the expected coverage proportions and Monte Carlo simulations, given $r_e$ and $r_c$, across $10,000$ trials.}\label{fig:demo_circle_case}
\end{figure}

For the Monte Carlo setup, we first fix the radius of the environment $r_e \in \{0.1, 1, 1.5, 2, 2.5\}$. For each fixed environment radius $r_e$, we sample $10,000$ random straight-line trajectories per $r_c$ coverage area radius, where $r_c \in \{kr_e/5\}_{k=1}^{k=5}$. The sampling method for Case (i) utilizes uniformly distributed endpoints, whereas for Case (ii) we use uniformly distributed midpoints. We compute the empirical mean of the coverage proportions across $10,000$ trials, as well as the standard deviation, and plot them in Figure \ref{fig:demo_circle_case}, overlaid with the direct evaluations of \eqref{eq:sol_case1} and \eqref{eq:sol_case2}. We observe a good match between our analytical expressions for the expected coverage proportions and the Monte Carlo results.

% In this section, we use Monte-Carlo (MC) method to generate samples and verify whether the expectation of these samples are equal to the analytical result. Comparisons and sensitivity analysis are made with varied hyperparameters with standard $10,000$ sample times for all cases. 

%  When the environment is defined as a circle with radius $R_e\in [0.5,1,1.5,2,2.5]$, the corresponding coverage area's radius are defined $R_c=[\frac{R_e}{5},\frac{2R_e}{5},\frac{3R_e}{5},\frac{4R_e}{5},R_e]$ over the domain, Figure \ref{fig:demo_circle_case}(a),(b) demonstrate the comparison of these $2$ cases' between analytical formation and MC, respectively.

% As we seem, all standard deviations are normalized, while the analytical and MC pairs are overlapping in the range of $E[P]$, which verified that $E[P]$ will increase until $1$ when $R_c\Rightarrow R_e$. 

\subsection{Numerical differences in expected coverage proportions} \label{ssec:num_diff_ECP}

\begin{figure}[!htbp]
	\centering
	\includegraphics[width=0.75\linewidth]{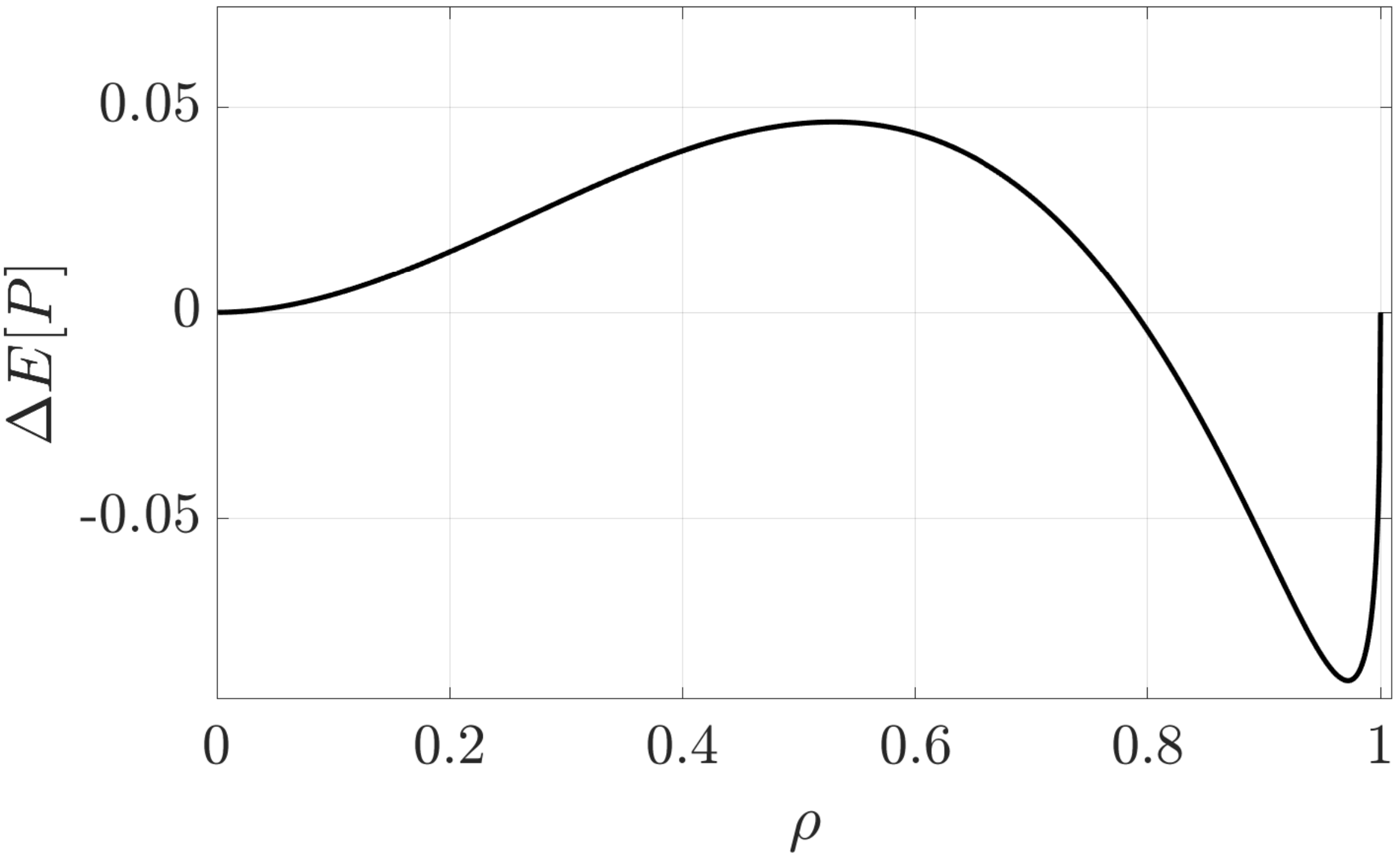}
	\caption{Numerical difference between the expected coverage proportion from Case (i) versus Case (ii), plotted against $\rho = r_c/r_e$.}
	\label{fig:sol_error_comp}
\end{figure}

Recall previously the discussion regarding the differences between Case (i) and (ii) in terms of the midpoint distributions within $\mathcal{D}_c$. However, as $r_c$ varies between $0$ and $r_e$, it is not clear what is the numerical difference between Case (i) and (ii). We would like to better understand when, e.g., the expected coverage proportion for Case (i) is \emph{greater} than Case (ii), for the same $r_e$ and $r_c$ values. We first note that this difference does not appear to depend on $r_e$ -- hence, we fix $r_e = 1$, and vary $\rho=r_c/r_e \in [0, 1]$. We plot the difference $\Delta\mathbb{E}[P] = \mathbb{E}\left[P_\mathrm{UDE}\right] - \mathbb{E}\left[P_\mathrm{UDM}\right]$ versus $\rho$ in Figure \ref{fig:sol_error_comp}.

We note that interestingly the expected coverage proportion for Case (i) is larger for a portion of $\rho$ values, then the expected coverage proportion for Case (ii) becomes larger past approximately $\rho = 0.79$ (recalling that $r_e$ is fixed at $1$). We observe two points where the difference between the two cases is maximized, first at approximately $\rho = 0.5$ (i.e., the environment radius is twice as big as the coverage radius), and when $\rho \approx 0.97$ (i.e., when the coverage area is almost as large as the environment). At the first extremum ($\rho \approx 0.5$), generating straight-line trajectories assuming uniformly distributed endpoints produce larger coverage proportions in expectation. At the second extremum ($\rho \approx 0.97$), generation via uniformly distributed midpoints produces larger expected coverage proportions. At the trivial cases when $\rho=0$ (no coverage area) and $\rho=1$ (coverage area matches the environment), the difference $\Delta\mathbb{E}[P]$ between the two expressions is 0, as expected.

\section{Urban Area Simulation-Based Case Studies}  \label{sec:urban_cases}

\subsection{Data set description}\label{sec:ideal_data_urban}

% source from: https://docs.google.com/spreadsheets/d/1LSQuQnaX4LyJ02SWU3rWEbEJsCSAst_jrgLVxvBFRYg/edit#gid=803486924
% \begin{table*}
%   \centering
%   \caption{Environment Comparison}
%     \begin{tabular}{cccccc}
%     \hline
%     \multirow{1}{*}{\textbf{City}} & \multicolumn{1}{c}{*Pop. Per Sqr Mile}     &  \multicolumn{1}{c}{**Stores Per Sqr Kilometer} &  \multirow{1}{*}{**Bldg. Per Sqr Mile} &  \multirow{1}{*}{**Mean Bldg. H.t. [m]} &  \multirow{1}{*}{**Std Bldg. H.t. [m]}   \\
%     \hline
%  \multirow{1}{*}{\textbf{SF}} & 7193 &1.30868 &1437 &7.85 &4.62\\
%     % \hline
%  \multirow{1}{*}{\textbf{NYC}} & 11314 &1.07690 &1830 &-- &--\\
%      % \hline
%  \multirow{1}{*}{\textbf{LA}} & 3206 &0.37254 &1918 &-- &--\\
%     \hline
%      \multicolumn{6}{c}{*\hyperlink{https://www.census.gov/quickfacts/fact/table/losangelescitycalifornia,newyorkcitynewyork,NY,sanfranciscocitycalifornia,sanfranciscocountycalifornia/POP060220}{US Census}\qquad \qquad  **OSM Dataset Stats}\\
%      % \hline
%     \end{tabular}%
%   \label{tab:1_sheet2}%
% \end{table*}

%Using OpenStreetMap data sets organized by MyGeoData automated much of the setup process foundational to conducting this paper's real-world case studies. 
After obtaining labeled geolocation data sets of buildings, vendors, and customers via MyGeoData and OpenStreetMap, we simplified the layouts of the San Francisco occupancy maps, origin points, and destination points to enable path planning simulations. Expanding on the discussion in Section \ref{ssec:urban_area_method_data}, MyGeoData breaks down OpenStreetMap data into pre-defined themes (e.g., airports, banks, cafes), and extracts all theme-associated data from a pre-defined region of interest (ROI). For our case study, we choose the MyGeoData themes of residential land use, shopping centers and department stores, and buildings in order to simulate the customers, vendors, and buildings in San Francisco. 

\begin{figure}[!htbp]
	\centering
	\includegraphics[width=\linewidth]{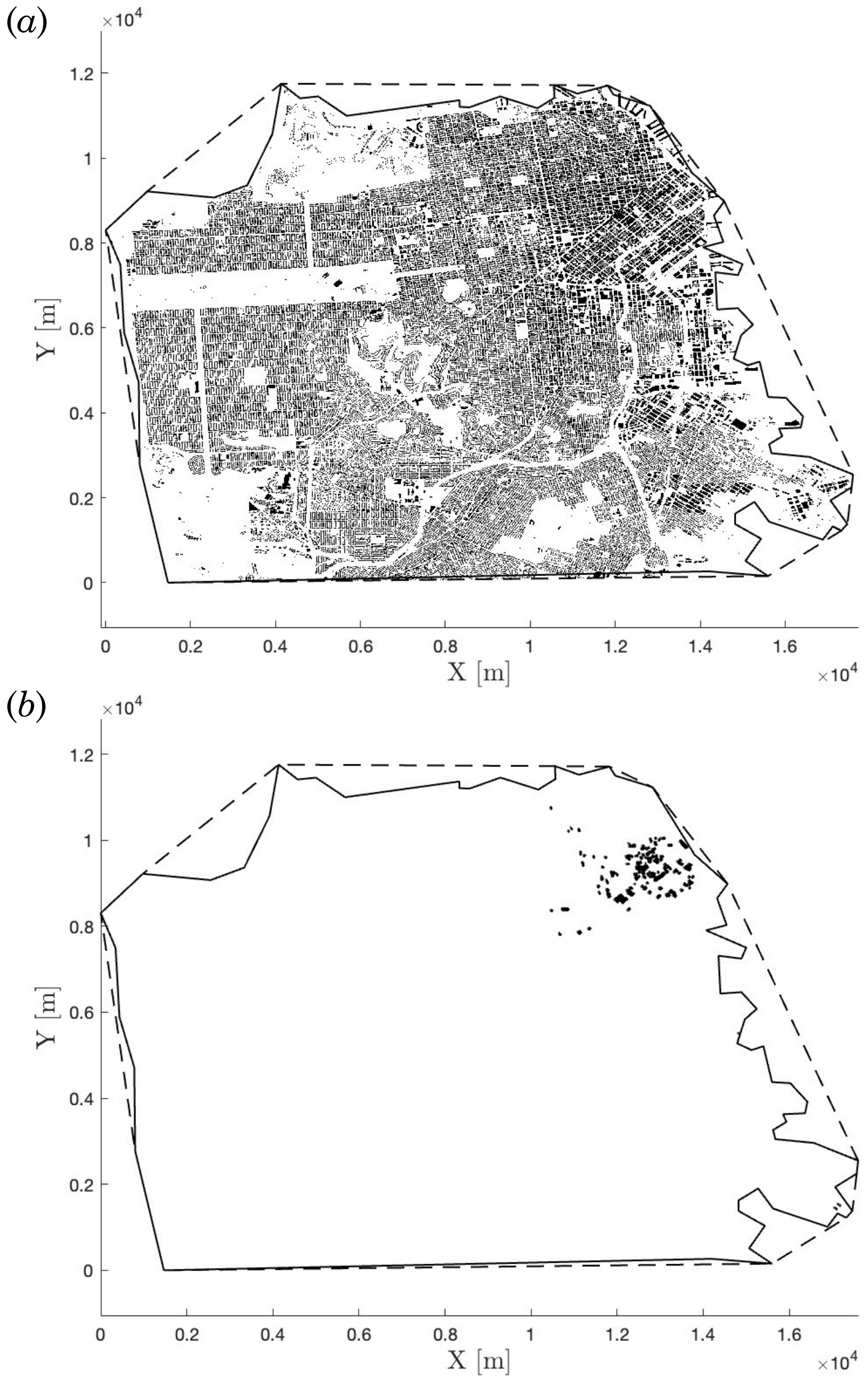}
	\caption{Building occupancy maps for (\emph{a}) San Francisco at 0 feet altitude and (\emph{b}) San Francisco at 200 feet altitude. Building occupancy at additional altitudes (i.e., 400 feet) used in the experiment is not shown for brevity.}
	\label{fig:occupancy_maps}
\end{figure}

The residential data sets highlight land containing residential dwellings, providing polygon geolocation coordinates for these areas. Customer locations are determined by randomly selecting the necessary number of geolocation coordinates within all 2D regions labeled as residential land to serve as possible drone destination locations. The data sets containing shopping centers and department stores feature land and hub centers associated with individual stores and supermarkets. These data sets provide analogous polygon and point geolocation coordinates for these commercial locations. One vendor location is assigned to each point and polygon to represent possible drone origin locations. Finally, the building data sets provide geolocation coordinates for the footprints of all buildings within the pre-defined ROI. The buildings serve as obstacles for the path planning algorithms when constructing the trajectory for an OD pair consisting of customers and vendors. We provide visualizations of the building occupancy at different altitudes in Figure \ref{fig:occupancy_maps}.

\begin{figure}[!htbp]
	\centering
	\includegraphics[width=\linewidth]{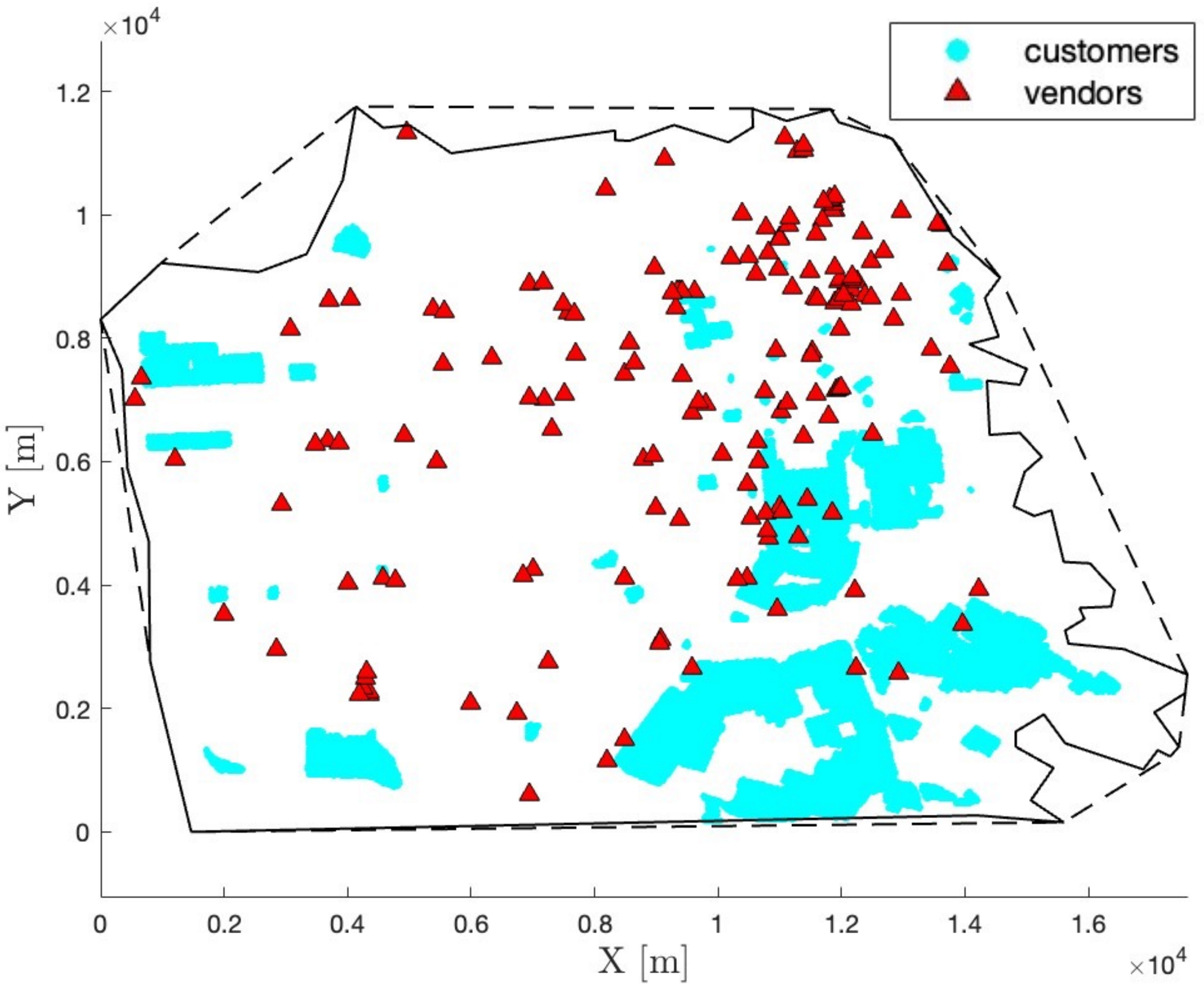}
	\caption{Locations of customers and vendors for San Francisco.}
	\label{fig:custvend_maps}
\end{figure}

We visualize the geographies of customers and vendors for each city in Figure \ref{fig:custvend_maps}. As can be seen in Figure \ref{fig:custvend_maps}, vendor and customer sites located outside of a given city's ROI boundaries are disregarded in the simulation. This is to adhere to city-specific customer population densities, as well as to ensure that UAS origins and destinations remain within the pre-defined ROI. We retain all buildings and potential Remote ID receiver locations within a convex polygon encompassing the city; we do this to ensure accurate representation of obstacles and potential coverage centers across all possible trajectories. Finally, in terms of case study environmental statistics, we note that San Francisco has a population per km\textsuperscript{2} of approximately 7,200 persons \cite{USCensus_Data}. Within San Francisco, there are 0.5 store sites per km\textsuperscript{2}, with approximately 555 buildings per km\textsuperscript{2} \cite{haklay2008openstreetmap,mygeodata}.

% Structural footprint polygons data from MyGeoData's  is converted to local vendor point coordinates. Residential area polygons with 2020 US Census population densities to random uniform distribution of customer point coordinates. 

% Building polygon information is used in creating the occupancy map and polygon map for the RRT* and Visibility Graph algorithms respectively. Creation of these maps involves algorithms removing merging adjacent polygons, deleting polygons inside other polygons, reducing the number of vertices, and adding geofences to accommodate RRT* cutting corners \cite{kim2022airspace}.

% Other items to discuss: Remote ID parameters (radius R1, ..., Rk), Remote ID distribution assumptions (spatial distribution of remote ID receivers),  Origin/destination selection (FIXED -- ONLY ONE WAY PER CITY), Base maps and alterations, etc. (different cities, NYC, SFO, etc.)

\subsection{Case study results}\label{sec:case_studies}

\begin{figure*}[!htbp]
	\centering
	\includegraphics[width=\linewidth]{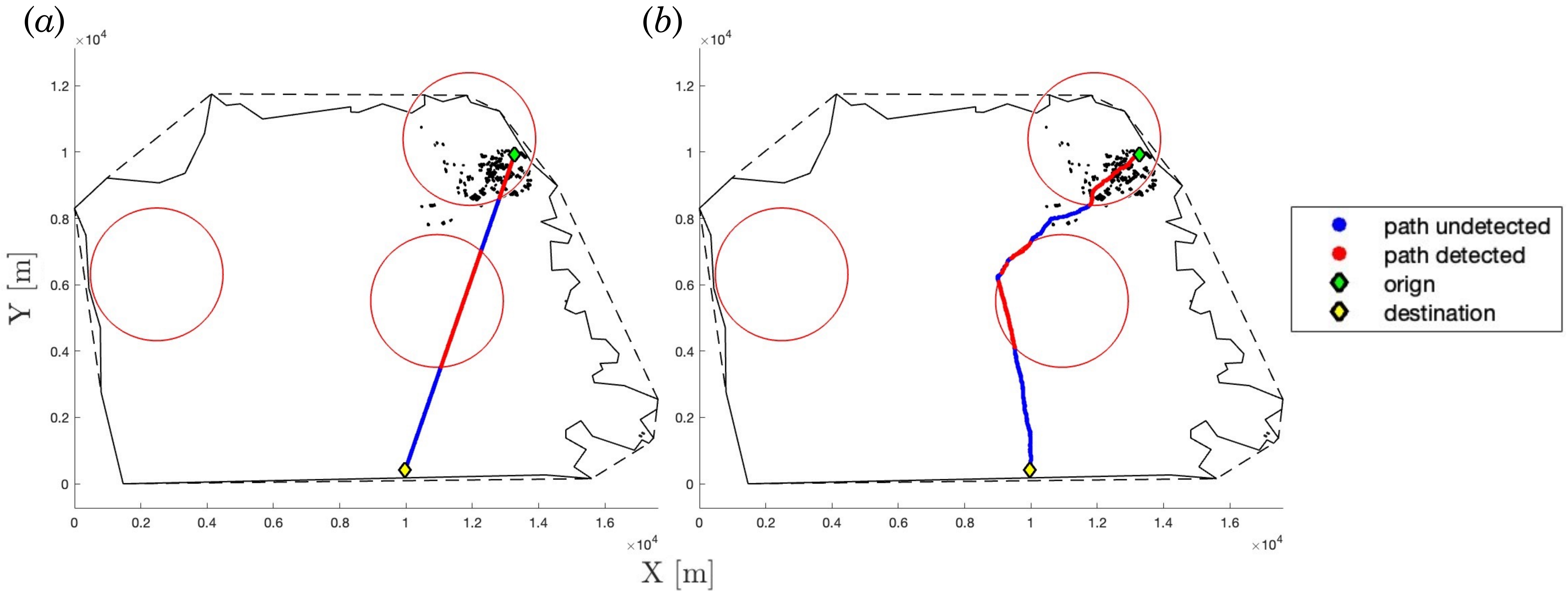}
	\caption{Sample OD drone paths at 200 feet cruising altitude, with 2 km radius Remote ID receivers overlaid, for San Francisco with (\emph{a}) \textsc{Slpp} and (\emph{b}) RRT\textsuperscript{*}.}
	\label{fig:paths_maps}
\end{figure*}

Prior to discussing the case study results, we provide visualizations of the case study setup in Figure \ref{fig:paths_maps}. Using data related to vendors and customers, we randomly select OD pairs for path planning. We also generate a random distribution of Remote ID receivers within the ROI. The coverage radius $r_c$ depends on the selected Remote ID receiver technology (e.g., Bluetooth, Wi-Fi), and coverage areas are visualized as red circles in Figure \ref{fig:paths_maps}. The coverage proportion of a single trajectory is given by the length of covered (i.e., detected, the red-colored portions of the trajectories in Figure \ref{fig:paths_maps}) trajectories divided by the total length of the trajectory. We continue sampling until we achieve convergence in terms of the average coverage proportion per scenario. Recall that a \emph{scenario} denotes a fixed altitude, a fixed Remote ID receiver technology, and iterating through a number of receivers (for \textsc{Slpp}). For RRT\textsuperscript{*}, due to its computational intensity compared to \textsc{Slpp}, the number of receivers is informed by the analogous scenario under \textsc{Slpp}. In addition, we only perform 1 trial of 200 randomly sampled trajectories for RRT\textsuperscript{*}; this is because we observe good convergence in terms of the coverage proportion, and it reduces the computation time required for conducting simulations involving RRT\textsuperscript{*}. Finally, we also examine the differences in average coverage proportions between the two path planning methods.

% Straight line results
% - generated 1000 paths for 20 fixed receiver distributions for each combination of radii, desired coverage proportion, and altitude) (reference table)
% - coverage uses discs to compute the total coverage proportion per path, (computing coverage proportion is invariant to cruise altitude) and straight line behavior assumes no obstacles, therefore the number of receivers needed for the straight line at each altitude is the same
% - for each combination/condition we are able to find the necessary number of receivers to achieve the desired coverage proportion within +-3.7%

% RRT*
% - RRT* coverage proportions at different altitudes were found to be larger t
% - 

% Notes from Testing:
% - found that raising the number of sensor distributions to 20 resulted in more consistent (less variance with the ) average trajectory coverage for the larger receiver ranges. To ensure average trajectories converged, 1000 paths per receiver distribution were generated

% - 

We list results for \textsc{Slpp} in Table \ref{tab:1_sheet3}. We present the number of Remote ID receivers needed to achieve three different desired coverage proportions, selecting between three different Remote ID receiver technologies (Bluetooth Legacy -- R250; Bluetooth Long Range -- R1000; Wi-Fi NAN/Beacon -- R2000). Note that for \textsc{Slpp}, we do not factor in the altitude as we are interested only in the straight-line path for an OD pair -- this is not the case for RRT\textsuperscript{*}. We observe that the required number of Remote ID receivers can vary drastically across different broadcast technologies, although the convergence in terms of average coverage proportions is good compared to the desired average coverage proportion. 

Results for RRT\textsuperscript{*} are listed in Table \ref{tab:2_sheet3}. Recall that for RRT\textsuperscript{*}, we fix the number of Remote ID receivers used to achieve a specific average coverage proportion for \textsc{Slpp}, and evaluate the converged average coverage proportion when using RRT\textsuperscript{*} as the path planning algorithm. We note some interesting differences, such as in the case between 200 versus 400 feet for the nominally 50\% coverage scenario: Even though \textsc{Slpp} achieved approximately 50\% coverage, using RRT\textsuperscript{*} results in a higher coverage proportion at higher altitudes. We also note particularly drastic cases of higher coverage under RRT\textsuperscript{*} for larger radii, as compared to \textsc{Slpp}.

\begin{table*}[ht]
  \centering
  \caption{Number of Remote ID receivers needed to achieve coverage proportions [\%] for \textsc{Slpp}, 20 trials of 1,000 trajectories each}
    \begin{tabular}{p{0.145\textwidth}<{\centering}p{0.111\textwidth}<{\centering}p{0.111\textwidth}<{\centering}p{0.111\textwidth}<{\centering}p{0.111\textwidth}<{\centering}p{0.111\textwidth}<{\centering}p{0.111\textwidth}<{\centering}}
    % \begin{tabular}{ccccccc}
    \hline
    \multirow{2}{*}{} 
      & \multicolumn{3}{c}{\textbf{Number of Remote ID Receivers}} 
      
      & \multicolumn{3}{c}{\textbf{Convergence Values [\%]}} \\ 
      &  $\mathbf{50\%}$ & $\mathbf{75\%}$ & $\mathbf{95\%}$ &  $\mathbf{50\%}$  & $\mathbf{75\%}$ & $\mathbf{95\%}$  \\\hline
     \multirow{1}{*}{\textbf{R250}} & 525 & 1300 & 5000 & 51.5 & 76.5 & 94.3\\
    \hline
     \multirow{1}{*}{\textbf{R1000}} & 30 & 65 & 160 & 51.6 & 74.8 & 94.7\\
    \hline
     \multirow{1}{*}{\textbf{R2000}}& 8 & 15 & 35 & 53.7 & 75.5 & 93.4 \\
    \hline
    \end{tabular}%
  \label{tab:1_sheet3}%
\end{table*}

\begin{table*}[ht]
  \centering
  \caption{Converged coverage proportions [\%] for RRT\textsuperscript{*} after 200 randomly sampled trajectories, given (fixed) number of Remote ID receivers required for 50\%, 75\%, and 95\% coverage under \textsc{Slpp}}
    \begin{tabular}{p{0.1\textwidth}<{\centering}p{0.0752\textwidth}<{\centering}p{0.0752\textwidth}<{\centering}p{0.0752\textwidth}<{\centering}p{0.0752\textwidth}<{\centering}p{0.0752\textwidth}<{\centering}p{0.06\textwidth}<{\centering}p{0.0752\textwidth}<{\centering}p{0.0752\textwidth}<{\centering}p{0.0752\textwidth}<{\centering}}
    % \begin{tabular}{ccccccc}
    \hline
    \multicolumn{10}{c}{\textbf{Convergence Values} [\%]} \\
    \multirow{2}{*}
    {} & \multicolumn{3}{c}{\textbf{R250}}       
      & \multicolumn{3}{c}{\textbf{R1000}} & \multicolumn{3}{c}{\textbf{R2000}}\\ 
     \textbf{Altitude} [ft.] &  $\mathbf{50\%}$ & $\mathbf{75\%}$ & $\mathbf{95\%}$ &  $\mathbf{50\%}$ & $\mathbf{75\%}$ & $\mathbf{95\%}$ & $\mathbf{50\%}$ & $\mathbf{75\%}$ & $\mathbf{95\%}$  \\\hline
     \multirow{1}{*} 200	& 54.7 & 75.4 & 94.1 & 48.3 & 71.3 & 92.5 & 71.9 & 82.1 & 91.3\\
    \hline
     \multirow{1}{*} 400 & 49.1 & 71.1 & 93.2 & 61.2 & 79.3 & 96.7 & 71.2 & 80.0 & 98.4\\
    \hline
    \end{tabular}%
  \label{tab:2_sheet3}%
\end{table*}

\begin{figure}[!htbp]
	\centering
	\includegraphics[width=0.8\linewidth]{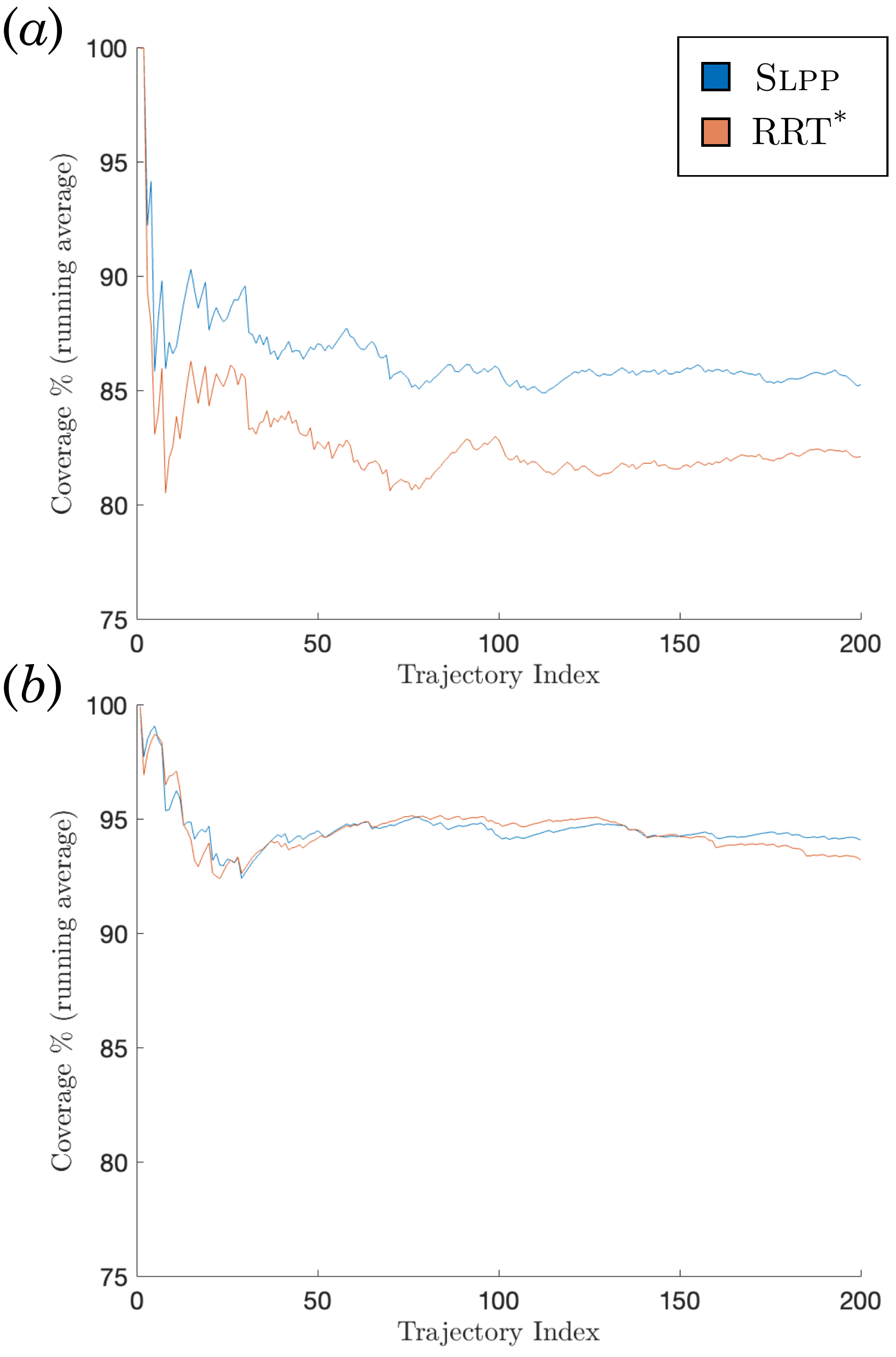}
	\caption{Running average of coverage proportions for \textsc{Slpp} and RRT\textsuperscript{*} across 200 sequential, independently sampled trajectories, under (\emph{a}) R2000 at 75\% desired coverage proportion, 200 feet cruising altitude; (\emph{b}) R250 at 95\% desired coverage proportion, 400 feet cruising altitude. Note that $y$-axis limits are identical for ease of comparison.}
	\label{fig:converge_plots}
\end{figure}

Finally, we briefly remark on observed convergences between \textsc{Slpp} and RRT\textsuperscript{*}. In Figure \ref{fig:converge_plots}, we plot 200 randomly sampled trajectories for two scenarios: R2000 (i.e., Wi-Fi NAN/Beacon) at 75\% desired coverage proportion, 200 feet cruising altitude in panel (\emph{a}), and R250 (i.e., Bluetooth Legacy) at 95\% desired coverage proportion, 400 feet cruising altitude in panel (\emph{b}). We observe that even after only 200 randomly generated trajectories, the running average of coverage proportions appear to stabilize. We note that in Figure \ref{fig:converge_plots}(\emph{a}), this was for one trial (out of 20) for \textsc{Slpp} -- the average across all trials is what is reported in Table \ref{tab:1_sheet3} as 75.5\%.

\subsection{Hybrid analysis outline} \label{ssec:integrated_analysis}

Recall from the idealized analysis that, assuming specific geometries for the environment $\mathcal{B}_e$ and coverage area $\mathcal{D}_c$, along with how trajectories between origins and destinations are generated, we can explicitly compute the expected coverage proportion. Given the computational intensity discussed previously in this section with respect to simulating individual trajectories, particularly if more sophisticated path planning algorithms are assumed to be used, a reasonable \enquote{hybrid} approach may be to use geographic partitioning, approximations, and repeated applications of, e.g., \eqref{eq:sol_case1} or \eqref{eq:sol_case2}. Using Figure \ref{fig:joint_analysis} as a visual guide, the outline for this hybrid approach is as follows:

\begin{figure}[!htbp]
	\centering
	\includegraphics[width=0.9\linewidth]{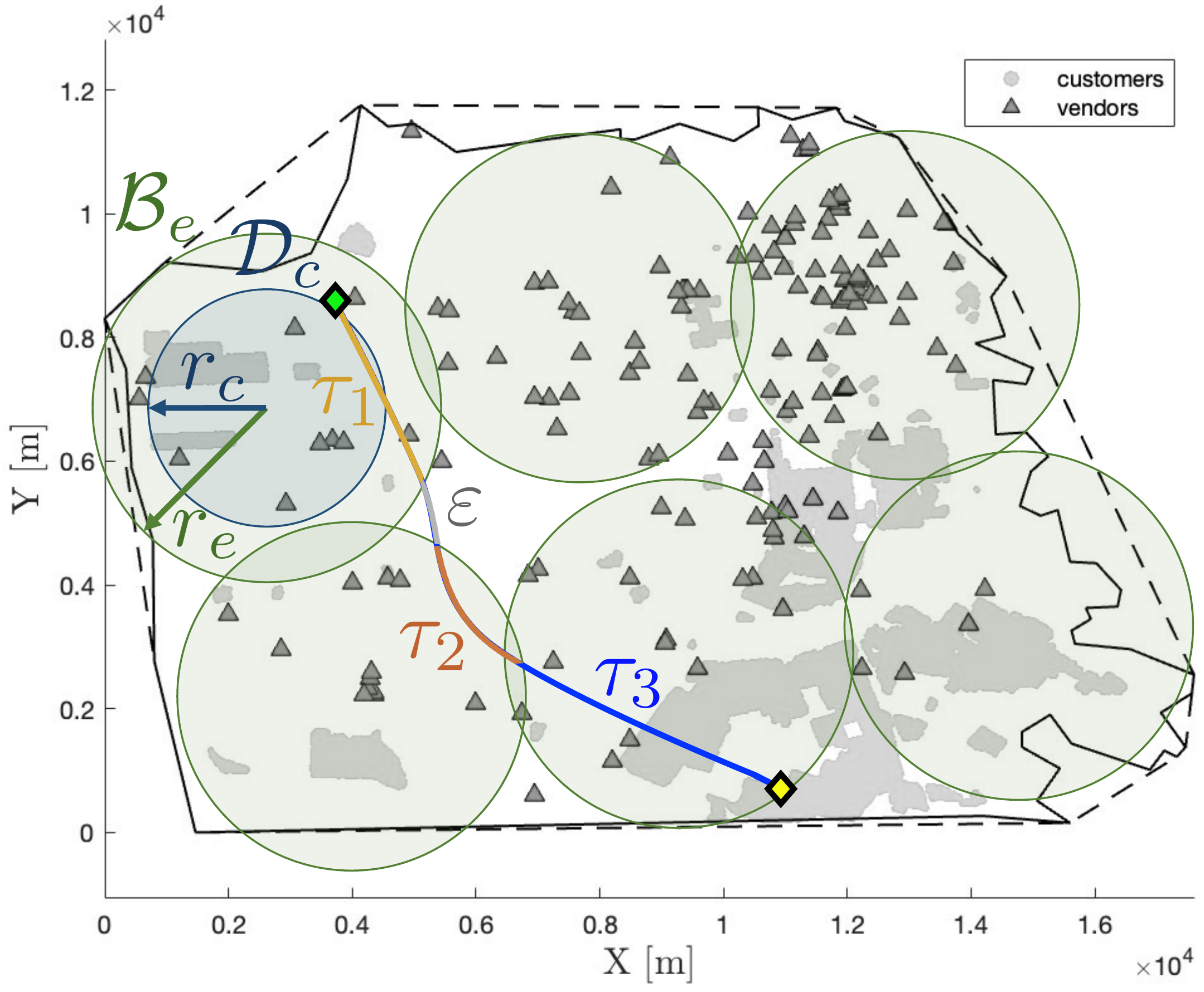}
	\caption{Notional overview figure of integrating the idealized coverage analysis from Section \ref{ssec:ideal_anl} with real-world geographies.}
	\label{fig:joint_analysis}
\end{figure}

\begin{enumerate}
    \item Select a desired $r_e$ for each individual $\mathcal{B}_e$ to be used in \emph{packing} the ROI. This selected $r_e$ should be greater than or equal to $r_c$ (fixed based on the Remote ID receiver technology of interest).
    \item Using a packing heuristic (e.g., \cite{leao2020_ipp}), pack the ROI with $K$ numbers of environments $\mathcal{B}_e^1, \hdots, \mathcal{B}_e^K$ each with its own coverage area $\mathcal{D}_c^1, \hdots, \mathcal{D}_c^K$. Note that $\mathcal{B}_e^i$ and $\mathcal{D}_c^i$ are centered at $\left(x_0^i, y_0^i\right)$ for $i = 1, \hdots, K$.
    \item Given a trajectory from an origin to a destination, decompose the trajectory into $\bigcup_i \tau_i$ where each $\tau_i$ is associated with a specific environment (and coverage area) $\{\mathcal{B}_e^i, \mathcal{D}_c^i\}$. 
    \item Compute the average coverage proportion expected per trajectory segment $\tau_i$ analytically via \eqref{eq:sol_case1} or \eqref{eq:sol_case2}, and report the final coverage proportions averaged across all possible trajectories in the ROI, along with approximation errors $\varepsilon$ as not all portions of the trajectory may be covered by an environment $\mathcal{B}_e^1, \hdots, \mathcal{B}_e^K$.
\end{enumerate}

\noindent
This approach of partitioning the trajectories per environment aligns most closely with the case of uniformly distributed endpoints, assuming that OD pairs are randomly generated with no consideration of $\mathcal{B}_e^i$ boundaries. Future analysis will be needed to determine exactly how the random process of OD generation maps to randomly sampled points on $\mathcal{B}_e^i$ boundaries. Additionally, an open question is how overlapping environments (as is shown in Figure \ref{fig:joint_analysis}) will impact Remote ID receiver coverage estimations. 
\section{Concluding Remarks}  \label{sec:conclusion}

Remote ID standards for UAM and AAM applications are safety-critical, and required for procedures such as counter-UAS operations, UAS traffic management (UTM), and low-altitude airspace management. In this work, we address the problem of trajectory coverage by Remote ID receivers in urban settings: Specifically, we assume a broadcast-receive architecture for Remote ID, and conduct (1) an idealized analysis of expected coverage proportions, as well as (2) a simulation-based case study with realistic urban geographies and path planning techniques. Under simplified geometries, we derived explicit equations for the expected coverage proportion, given the coverage radius of a Remote ID receiver. For the urban case studies, we explored the number of Remote ID receivers needed to achieve specific average coverage proportions in San Francisco. In designing the San Francisco case study, we considered different path planning assumptions, vendor and customer densities, Remote ID receiver distributions and broadcast technologies, as well as cruise altitude. The results and models from our idealized and simulation-based urban case study can be used by municipal authorities and AAM stakeholders for guidance when implementing future Remote ID systems for UAM and AAM applications.

\subsection{Limitations and future work}    \label{ssec:limitations_fw}

\begin{figure}[!htbp]
	\centering
	\includegraphics[width=0.75\linewidth]{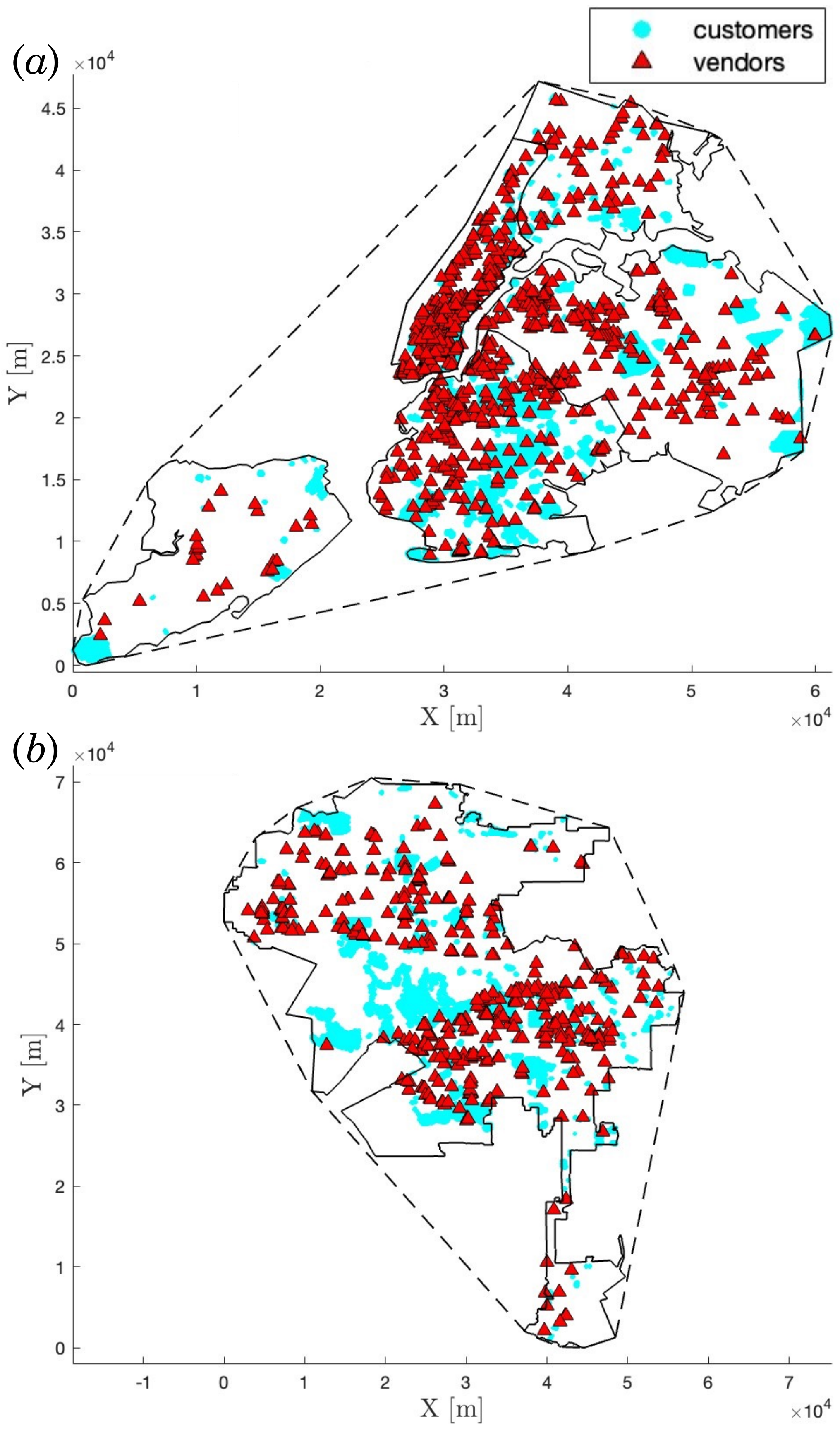}
	\caption{Locations of customers and vendors for (\emph{a}) New York City and (\emph{b}) Los Angeles.}
	\label{fig:custvend_maps_NYC_LA}
\end{figure}

For our idealized analysis, we assumed specific geometries and overlaps between the coverage area and the environment. Furthermore, for both the idealized analysis and the urban case study, we made simplifying assumptions regarding drone dynamics as well as the reception capabilities of the Remote ID receiver (e.g., \cite{kuroda2020_uascoms} focuses on communication bandwidth in Remote ID setups). Readily-available extensions include performing the coverage characterization using different city geographies, such as New York City and Los Angeles (see Figure \ref{fig:custvend_maps_NYC_LA}). Future work includes relaxing assumptions that leads to generalizing our idealized analysis (e.g., not requiring the Remote ID receiver to be centered at the environment's center), incorporating random dropout and communication errors, along with field testing to validate our results using real drones and physical Remote ID setups.

% \noindent
% \textbf{Limitations and Future Work:} 

% \noindent
% \textbf{Limitations:}

% \noindent
% \textbf{Future work:}

% VI. Concluding Remarks % ML
%     1. Limitations of Work
%         - Simplifying assumptions made in idealized coverage analysis
%         - Simplifying assumptions regarding drone dynamics/simulation/etc.
%     2. Future lines of research/future work
%         - More realistic implementations, etc.
%         - Link with UAV privacy work, which assumed 100\% coverage, we will show some other proportions, so how do you bridge the gap (e.g., predicting rest of trajectory, how accurately can you do so, etc.)
%         - Thinking about more path planning/trajectory generation
%             - Kinematic feasibility OR
%             - Easier to compute, etc.

%             - Field testing!
%\input{ridcov_atm2023/chapters/7_appendix.tex}

\bibliographystyle{IEEEtran} % use IEEEtran.bst style
\bibliography{main.bib}

\begin{thebibliography}{10}
\providecommand{\url}[1]{#1}
\csname url@samestyle\endcsname
\providecommand{\newblock}{\relax}
\providecommand{\bibinfo}[2]{#2}
\providecommand{\BIBentrySTDinterwordspacing}{\spaceskip=0pt\relax}
\providecommand{\BIBentryALTinterwordstretchfactor}{4}
\providecommand{\BIBentryALTinterwordspacing}{\spaceskip=\fontdimen2\font plus
\BIBentryALTinterwordstretchfactor\fontdimen3\font minus
  \fontdimen4\font\relax}
\providecommand{\BIBforeignlanguage}[2]{{%
\expandafter\ifx\csname l@#1\endcsname\relax
\typeout{** WARNING: IEEEtran.bst: No hyphenation pattern has been}%
\typeout{** loaded for the language `#1'. Using the pattern for}%
\typeout{** the default language instead.}%
\else
\language=\csname l@#1\endcsname
\fi
#2}}
\providecommand{\BIBdecl}{\relax}
\BIBdecl

\bibitem{goyal2021_AAM}
R.~Goyal, C.~Reiche, C.~Fernando, and A.~Cohen, ``{Advanced Air Mobility:
  Demand Analysis and Market Potential of the Airport Shuttle and Air Taxi
  Markets},'' \emph{Sustainability}, vol.~13, no.~13, 2021.

\bibitem{NASAAAM_2022}
{National Aeronautics and Space Administration}, ``{Advanced Air Mobility},''
  2022, accessed: January 2023.

\bibitem{cohen2021_AAMUseCases}
A.~P. Cohen, S.~A. Shaheen, and E.~M. Farrar, ``{Urban Air Mobility: History,
  Ecosystem, Market Potential, and Challenges},'' \emph{IEEE Transactions on
  Intelligent Transportation Systems}, vol.~22, no.~9, pp. 6074--6087, 2021.

\bibitem{sigari2021_AAMM}
C.~Sigari and P.~Biberthaler, ``{Medical drones: Disruptive technology makes
  the future happen},'' \emph{Der Unfallchirurg}, vol. 124, no.~12, pp.
  974--976, 2021.

\bibitem{cohen2021_UC}
A.~Cohen and S.~Shaheen, ``{Urban Air Mobility: Opportunities and Obstacles},''
  in \emph{International Encyclopedia of Transportation}, R.~Vickerman,
  Ed.\hskip 1em plus 0.5em minus 0.4em\relax Oxford: Elsevier, 2021, pp.
  702--709.

\bibitem{ellis2021_IASMS}
K.~K. Ellis, P.~Krois, J.~Koelling, L.~J. Prinzel, M.~Davies, and R.~Mah, ``{A
  Concept of Operations (ConOps) of an In-time Aviation Safety Management
  System (IASMS) for Advanced Air Mobility (AAM)},'' in \emph{AIAA Scitech 2021
  Forum}, 2021.

\bibitem{FAA_RemoteID_2021}
{Federal Aviation Administration}, ``{UAS Remote Identification Overview},''
  2021, accessed: January 2022.

\bibitem{RID_FR}
{Federal Aviation Administration (FAA), Department of Transportation (DOT)},
  ``{Remote Identification of Unmanned Aircraft},'' \emph{Federal Register},
  2020.

\bibitem{EASA_FR}
{European Union Aviation Safety Agency}, ``{Easy Access Rules for Unmanned
  Aircraft Systems (Regulations (EU) 2019/947 and (EU) 2019/945)},''
  \emph{European Union}, 2020.

\bibitem{RID_Court}
{United States Court of Appeals for the District of Columbia}, \emph{{Tyler
  Brennan and Racedayquads LLC, v. Stephen Dickson, Administrator and Federal
  Aviation Administration}}.\hskip 1em plus 0.5em minus 0.4em\relax {United
  States Court of Appeals for the District of Columbia}, 2022.

\bibitem{astm2022_remoteidstandard}
{ASTM International}, ``{Standard Specification for Remote ID and Tracking},''
  {ASTM International}, Tech. Rep. ASTM F3411-22a, 2022.

\bibitem{ding2022routing}
G.~Ding, A.~Berke, K.~Gopalakrishnan, K.~Degue, H.~Balakrishnan, and M.~Z. Li,
  ``Routing with privacy for drone package delivery systems,'' in
  \emph{International Conference on Research in Air Transportation}, June 2022.

\bibitem{woodcock2022_noise}
J.~Woodcock, S.~Hasan, J.~G. Paje, R.~Biziorek, P.~Henning, A.~Guthrie,
  V.~Jurdic, A.~L. Maldonado, and D.~Hiller, ``{Development of a noise-based
  route optioneering tool for advanced air mobility (AAM) vehicles},'' in
  \emph{28th AIAA/CEAS Aeroacoustics 2022 Conference}, 2022.

\bibitem{wang2021_cuas}
J.~Wang, Y.~Liu, and H.~Song, ``{Counter-Unmanned Aircraft System(s) (C-UAS):
  State of the Art, Challenges, and Future Trends},'' \emph{IEEE Aerospace and
  Electronic Systems Magazine}, vol.~36, no.~3, pp. 4--29, 2021.

\bibitem{young2020_faatest}
R.~Young, ``{UTM Evolution Into the 2020S – New York as a Case Study},'' in
  \emph{2020 Integrated Communications Navigation and Surveillance Conference
  (ICNS)}, 2020.

\bibitem{belwafi2022_ridts}
K.~Belwafi, R.~Alkadi, S.~A. Alameri, H.~A. Hamadi, and A.~Shoufan, ``{Unmanned
  Aerial Vehicles’ Remote Identification: A Tutorial and Survey},''
  \emph{IEEE Access}, vol.~10, 2022.

\bibitem{mujumdar2021_lora}
O.~Mujumdar, H.~Celebi, I.~Guvenc, M.~Sichitiu, S.~Hwang, and K.-M. Kang,
  ``{Use of LoRa for UAV Remote ID with Multi-User Interference and Different
  Spreading Factors},'' in \emph{2021 IEEE 93rd Vehicular Technology Conference
  (VTC2021-Spring)}, 2021, pp. 1--7.

\bibitem{wu2014_ltenetworks}
D.~Wu, L.~Zhou, Y.~Cai, R.~Q. Hu, and Y.~Qian, ``{The role of mobility for D2D
  communications in LTE-advanced networks: energy vs. bandwidth efficiency},''
  \emph{IEEE Wireless Communications}, vol.~21, no.~2, pp. 66--71, 2014.

\bibitem{kuroda2020_uascoms}
V.~Kuroda, M.~Egorov, S.~Munn, and A.~Evans, ``{Unlicensed Technology
  Assessment for Uas Communications},'' in \emph{2020 Integrated Communications
  Navigation and Surveillance Conference (ICNS)}, 2020.

\bibitem{megueridichian2001_coverage}
S.~Meguerdichian, F.~Koushanfar, M.~Potkonjak, and M.~Srivastava, ``{Coverage
  problems in wireless ad-hoc sensor networks},'' in \emph{Proceedings IEEE
  INFOCOM 2001. Conference on Computer Communications. Twentieth Annual Joint
  Conference of the IEEE Computer and Communications Society (Cat.
  No.01CH37213)}, vol.~3, 2001, pp. 1380--1387.

\bibitem{huang2003_coverage}
C.-F. Huang and Y.-C. Tseng, ``{The Coverage Problem in a Wireless Sensor
  Network},'' in \emph{Proceedings of the 2nd ACM International Conference on
  Wireless Sensor Networks and Applications}, 2003, p. 115–121.

\bibitem{paull2014_uavcoverage}
L.~Paull, C.~Thibault, A.~Nagaty, M.~Seto, and H.~Li, ``{Sensor-Driven Area
  Coverage for an Autonomous Fixed-Wing Unmanned Aerial Vehicle},'' \emph{IEEE
  Transactions on Cybernetics}, vol.~44, no.~9, pp. 1605--1618, 2014.

\bibitem{wang2011_coveragesurvey}
B.~Wang, ``{Coverage Problems in Sensor Networks: A Survey},'' \emph{ACM
  Comput. Surv.}, vol.~43, no.~4, 2011.

\bibitem{luo2019_tfe}
X.~Luo, B.~Liu, P.~J. Jin, Y.~Cao, and W.~Hu, ``{Arterial Traffic Flow
  Estimation Based on Vehicle-to-Cloud Vehicle Trajectory Data Considering
  Multi-Intersection Interaction and Coordination},'' \emph{Transportation
  Research Record}, vol. 2673, no.~6, pp. 68--83, 2019.

\bibitem{tong2021_camera}
P.~Tong, M.~Li, M.~Li, J.~Huang, and X.~Hua, ``{Large-Scale Vehicle Trajectory
  Reconstruction with Camera Sensing Network},'' in \emph{Proceedings of the
  27th Annual International Conference on Mobile Computing and Networking},
  2021, p. 188–200.

\bibitem{marinoff1994_bp}
L.~Marinoff, ``A resolution of bertrand's paradox,'' \emph{Philosophy of
  Science}, vol.~61, no.~1, pp. 1--24, 1994.

\bibitem{leonard2021survey}
C.~Leonard, L.~A. Garrow, and J.~Newman, ``{A Survey to Model Demand for eVTOL
  Trips to Airports},'' in \emph{AIAA AVIATION 2021 FORUM}, 2021, p. 3180.

\bibitem{kim2022airspace}
J.~Kim and E.~Atkins, ``{Airspace Geofencing and Flight Planning for
  Low-Altitude, Urban, Small Unmanned Aircraft Systems},'' \emph{Applied
  Sciences}, vol.~12, no.~2, p. 576, 2022.

\bibitem{wooten2019economic}
B.~Wooten and E.~Alvarez, ``{Economic Analysis of UAV Application for USPS Mail
  Delivery in Los Angeles County},'' \emph{2019 RSCA Oral Presentation
  Sessions}, 2019.

\bibitem{jayaraman2022rooftop}
J.~Jayaraman, V.~R. Balu, S.~Bregni, D.~Scazzoli, and M.~Magarini, ``{Rooftop
  Relay Nodes to Enhance URLLC in UAV-Assisted Cellular Networks},'' in
  \emph{ICC 2022-IEEE International Conference on Communications}.\hskip 1em
  plus 0.5em minus 0.4em\relax IEEE, 2022, pp. 733--738.

\bibitem{haklay2008openstreetmap}
M.~Haklay and P.~Weber, ``{Openstreetmap: User-generated street maps},''
  \emph{IEEE Pervasive computing}, vol.~7, no.~4, pp. 12--18, 2008.

\bibitem{mygeodata}
\BIBentryALTinterwordspacing
``{MyGeoData Cloud}.'' [Online]. Available: \url{https://mygeodata.cloud/}
\BIBentrySTDinterwordspacing

\bibitem{orlik2014current}
A.~Orlik and L.~Orlikova, ``{Current trends in formats and coordinate
  transformations of geospatial data—based on MyGeoData Converter},''
  \emph{Central European Journal of Geosciences}, vol.~6, pp. 354--362, 2014.

\bibitem{ASD-STAN_EuropeanRemoteID}
{Aerospace and Defense Industries Association of Europe - Standardization},
  ``{Direct Remote ID: Introduction to the European UAS Digital Remote ID
  Technical Standard},'' {ASD-STAN}, Tech. Rep., 2021.

\bibitem{chin2021_utm}
{Chin, Christopher and Gopalakrishnan, Karthik and Egorov, Maxim and Evans,
  Antony and Balakrishnan, Hamsa}, ``Efficiency and fairness in unmanned air
  traffic flow management,'' \emph{IEEE Transactions on Intelligent
  Transportation Systems}, vol.~22, no.~9, pp. 5939--5951, 2021.

\bibitem{karaman2011_pp}
{Karaman, Sertac and Frazzoli, Emilio}, ``Sampling-based algorithms for optimal
  motion planning,'' 2011.

\bibitem{zammit2018_rrt}
{Christian Zammit and Erik-Jan Van Kampen}, ``Comparison between a* and rrt
  algorithms for uav path planning,'' in \emph{2018 AIAA Guidance, Navigation,
  and Control Conference}, 2018.

\bibitem{FAA2021_RSKY}
{Federal Aviation Administration}, ``{Airspace 101 - Rules of the Sky},'' 2021,
  accessed: January 2023.

\bibitem{USCensus_Data}
{United States Census Bureau}, ``{QuickFacts - United States},'' 2022,
  accessed: February 2023.

\bibitem{leao2020_ipp}
A.~A. Leao, F.~M. Toledo, J.~F. Oliveira, M.~A. Carravilla, and
  R.~Alvarez-Vald\'{e}s, ``{Irregular packing problems: A review of
  mathematical models},'' \emph{European Journal of Operational Research}, vol.
  282, no.~3, pp. 803--822, 2020.

\end{thebibliography}

\section*{Author Biographies}

\small{ {\bf Hejun Huang} is a PhD student in the Department of Aerospace Engineering at the University of Michigan, Ann Arbor. He earned his MSc in Mechanical and Automation Engineering from The Chinese University of Hong Kong in 2020 and his BSE in Mechatronic Engineering from the North China Electric Power University in 2019. Prior to Ann Arbor, he was a research assistant at The Chinese University of Hong Kong. His research interests include air traffic systems, networked systems, and stability in large-scale systems with transitions. \vspace{1 mm}}

\small{ {\bf Billy Mazotti} is a Master's student in the Department of Aerospace Engineering at the University of Michigan, Ann Arbor. He earned his BS in Mechanical Engineering and BS Aerospace Science and Engineering from the University of California, Davis in 2021. His research interests include include robotics for space applications, computer vision in dynamic and unstructured environments, and trajectory optimization. \vspace{1 mm}}

\small{ {\bf Joseph Kim} is a PhD Candidate in the Robotics Department at the University of Michigan, where he researches geofencing algorithms, contingency management, and network management for UTM systems. He holds a BSE in Aerospace Engineering from the University of Texas at Austin, and a MSE in Aerospace Engineering (Autonomous Systems and Control) from the University of Michigan. His research interests are safe and efficient autonomous air traffic management with potential applications in drones and Urban Air Mobility. \vspace{1 mm}}

\small{ {\bf Max Z. Li} is an Assistant Professor in the Department of Aerospace Engineering and Department of Industrial and Operations Engineering at the University of Michigan, Ann Arbor. Max received his PhD in Aerospace Engineering from the Massachusetts Institute of Technology in 2021. He earned his MSE in Systems Engineering and BSE in Electrical Engineering and Mathematics, both from the University of Pennsylvania, in 2018. His research and teaching interests include air transportation systems, airport and airline operations, UAM/AAM, networked systems, as well as optimization and control.}

\section{Appendix}  \label{sec:appendix}

\flushend
\end{document}